\documentclass[a4paper,11pt]{article}
\usepackage{float}
\usepackage{jcappub}
\usepackage{afterpage}
\usepackage[utf8]{inputenc}
\usepackage{epsf}
\usepackage{graphicx} 
\usepackage{amssymb}
\usepackage{dcolumn}
\usepackage{amsmath}
\usepackage{hyperref}
\usepackage{multirow}
\usepackage{booktabs}
\usepackage{siunitx}
\usepackage{color}
\usepackage{subcaption}

\def\thetaB{\mbox{\boldmath$\theta$}}

\def\lsim{~\rlap{$<$}{\lower 1.0ex\hbox{$\sim$}}}
\def\gsim{~\rlap{$>$}{\lower 1.0ex\hbox{$\sim$}}}

\def\red{\color{black}}

\def\xB{\mbox{\boldmath$x$}}

\title{MassiveNuS: Cosmological Massive Neutrino Simulations}

\author[a,1,2]{Jia Liu, \note{Corresponding author.}\note{NSF Astronomy and Astrophysics Postdoctoral Fellow.}}
\emailAdd{jia@astro.princeton.edu}
\author[b,c]{Simeon Bird,}
\author[d]{Jos\'e Manuel Zorrilla Matilla,}
\author[d,e,f]{J. Colin Hill,}
\author[d,g]{Zolt\'an Haiman,}
\author[a]{Mathew S. Madhavacheril,}
\author[h]{Andrea Petri,}
\author[a,f]{and David N. Spergel}

\affiliation[a]{Department of Astrophysical Sciences, Princeton University, Princeton, NJ 08544, USA}
\affiliation[b]{Department of Physics and Astronomy, UC Riverside, Riverside, CA 92521, USA}
\affiliation[c]{Department of Physics and Astronomy, Johns Hopkins University, Baltimore, Maryland 21218, USA}
\affiliation[d]{Department of Astronomy, Columbia University, New York, NY 10027, USA} 
\affiliation[e]{Institute for Advanced Study, Princeton, NJ 08540, USA}
\affiliation[f]{Center for Computational Astrophysics, Flatiron Institute, New York, NY 10003, USA}
\affiliation[g]{Institute for Strings, Cosmology, and Astroparticle Physics, Columbia University, New York, NY 10027, USA}
\affiliation[h]{Department of Physics, Columbia University, New York, NY 10027, USA} 

\abstract{The non-zero mass of neutrinos suppresses the growth of cosmic structure on small scales. Since the level of suppression depends on the sum of the masses of the three active neutrino species, 
the evolution of large-scale structure is a promising tool to constrain the total mass of neutrinos and possibly shed light on the mass hierarchy. In this work, we investigate these effects via a large suite of $N$-body simulations that include massive neutrinos using an analytic linear-response approximation: the Cosmological Massive Neutrino Simulations (MassiveNuS). The simulations include the effects of radiation on the background expansion, as well as the clustering of neutrinos in response to the nonlinear dark matter evolution. We allow three cosmological parameters to vary: the neutrino mass sum $M_\nu$ in the range of 0--0.6~eV, the total matter density $\Omega_m$, and the primordial power spectrum amplitude $A_s$. The rms density fluctuation in spheres of 8 comoving Mpc/$h$ ($\sigma_8$) is a derived parameter as a result. Our data products include $N$-body snapshots, halo catalogues, merger trees, ray-traced galaxy lensing convergence maps for four source redshift planes between $z_s$=1--2.5, and ray-traced cosmic microwave background lensing convergence maps. We describe the simulation procedures and code validation in this paper. The data are publicly available at {\red \url{http://columbialensing.org}}.}

\begin{document}
\maketitle

\section{Introduction}\label{sec:intro}

Neutrino masses ($m_\nu$) are now known to be non-zero via the discovery of oscillations between the flavor eigenstates~\cite{Becker-Szendy1992,Fukuda1998,Ahmed2004}, a discovery {\red of physics} beyond the standard model of particle physics. Currently, only the differences between the squared masses of the three neutrino species are known: $\Delta m_{21}^2 \equiv m_2^2-m_1^2=7.37\substack{+0.60 \\ -0.44}\times10^{-5}$~eV$^2$ 
and $|\Delta m^2| \equiv |m_3^2-(m_1^2+m_2^2)/2|=2.5\substack{+0.13 \\ -0.13}\times 10^{-3}$~eV$^2$
from oscillation experiments~\cite{Olive:2016xmw} (99.73\% CL, normal hierarchy). Because the sign of $\Delta m^2$ is unknown, there are two possible ways of ranking the three neutrino masses: the ``normal'' hierarchy where $m_1<m_2<m_3$ or the ``inverted'' hierarchy where $m_3<m_1<m_2$, with a minimum summed mass $M_\nu \equiv \sum_i m_i \approx 0.06$~eV and $M_\nu \approx 0.1$~eV, respectively. The sum of the neutrino masses is, however, hard to determine using particle physics experiments, {\red such as KATRIN~\cite{Wolf2010} and Project 8~\cite{Esfahani2017}, which are only sensitive to the mass of the lightest neutrino}.

Massive neutrinos change the expansion history and growth of structure in the universe~(for recent reviews, see~\cite{LesgourguesPastor2006,Wong2011}). First, neutrinos of mass less than a few eV remain relativistic and behave like radiation at the time of matter-radiation equality, but are non-relativistic and behave like matter now. Therefore, for a fixed present-day matter density, $\Omega_m$=$\Omega_{c}$+$\Omega_{b}$+$\Omega_\nu$, massive neutrinos move the matter-radiation equality epoch to a later time, hence delaying the onset of linear growth. Here, $\Omega_c$ is the density of the cold dark matter~(CDM) component, $\Omega_b$ is the baryon density, and $\Omega_\nu$ is the density of massive neutrinos, all in units of the critical density.

Second, neutrinos with small masses have large thermal velocities, and can stream out of CDM potential wells freely, suppressing the growth of structure below a certain scale. This scale is approximately equal to the Hubble radius when neutrinos are relativistic. When neutrinos become non-relativistic, at redshift $z_{\rm nr}\approx m_\nu c^2/ k_B T_{\nu,0}$, the scale is where gravitational instability overcomes the thermal pressure of neutrinos (i.e., the neutrino Jeans scale). Here, $c$ is the speed of light, $k_B$ is the Boltzmann constant, and $T_{\nu,0}=1.95$~K is the temperature of cosmic neutrinos today (assuming a standard thermal history). For $\approx 0.1$~eV neutrinos, for example, $z_{\rm nr}\approx600$ and the free-streaming scale is around 110 Mpc today.
Due to neutrino free-streaming, the suppression is more prominent at smaller scales, where nonlinear growth dominates. 

Because of these effects, observations of large-scale structure and its evolution can constrain the total mass of neutrinos. In the case of a mass sum $<0.1$~eV, such observations can also constrain the neutrino mass hierarchy.  Cosmological constraints on the neutrino mass sum ($M_\nu$) are quickly approaching the lower limit implied by neutrino oscillation data. For example, the Planck team obtained an upper limit of $M_\nu<0.23$~eV~\cite{planck2015xiii} (95\% CL) 
using primary cosmic microwave background (CMB) temperature data, combined with low-$\ell$ polarization, CMB lensing, type Ia supernovae~\cite{Betoule2014}, and baryon acoustic oscillation
measurements~\cite{Beutler2011,Anderson2014,Ross2015}. Ref.~\cite{Palanque-Delabrouille2015} found a tighter constraint of $M_\nu<0.15$~eV by combining Ly$\alpha$ forest data from the Baryon Oscillation Spectroscopic Survey of the Sloan Digital Sky Survey and 2013 Planck data~\cite{planck2013xvi}.
Data from upcoming large cosmological surveys, such as the Large Synoptic Survey Telescope~(LSST)\footnote{\url{http://www.lsst.org} }, Wide-Field Infrared Survey Telescope~(WFIRST)\footnote{\url{http://wfirst.gsfc.nasa.gov} }, Euclid\footnote{\url{http://sci.esa.int/euclid} }, Simons Observatory\footnote{\url{http://www.simonsobservatory.org} }, and CMB-S4\footnote{\url{https://cmb-s4.org}} will provide high-precision measurements of the matter density field. In order to obtain a significant detection, for which an uncertainty of $\sigma(M_{\nu}) \approx 0.01$~eV is desired, accurate modeling of neutrino effects on the matter density field is critical. 

On large scales, $k\lesssim$ 0.2~$h$/Mpc, where $k$ is the wavenumber in Fourier space, matter fluctuations evolve linearly or quasi-linearly. In this regime, the matter power spectrum $P(k)$ can be modeled analytically~(e.g.,~\cite{Bond1980,Zeldovich1980,Ma1994,Valdarnini1998}). On small scales, gravitational interactions lead to nonlinear evolution of perturbations. Therefore high-resolution $N$-body simulations must be utilized to model $P(k)$. The \texttt{Halofit} formalism~\cite{smith2003,takahashi2012,Bird2012,Mead2015,Mead2016,Casarini2009,Casarini2016}, based on the ``halo model''~\cite{PeacockSmith2000,Seljak2000} and calibrated against numerical simulations, has been widely used to provide a fitting function for the matter power spectrum in the nonlinear regime. Alternatively, a series of papers dubbed the ``Coyote Universe'' developed a prediction scheme called the ``emulator'', based on interpolating the measured $P(k)$ from a large number of simulations with various resolutions and cosmological models.  The emulator reached percent-level accuracy out to $k\approx 10$/Mpc and $z=4$ for $w$CDM cosmologies~\cite{heitmann2010i,heitmann2009ii,heitmann2010iii,heitmann2014e}. However, among all these efforts, only \cite{Bird2012} included massive neutrino {\red particles} in their simulation, and with only 3 different neutrino masses (0.15, 0.3 and 0.6~eV). Our work has a much denser sampling of neutrino masses, and hence is suitable to study the nonlinear effects of massive neutrinos as well as their degeneracy with other cosmological parameters.

In this paper, we present a suite of 101 $N$-body simulations---``MassiveNuS'' (Cosmological Massive Neutrino Simulations). The effect of massive neutrinos is included following \cite[][hereafter AB13, B17]{AlihaimoundBird2013, Bird2017}. Our method extends that of Ref.~\cite{Brandbyge2009}, who had the insight that neutrinos are still perturbative, even at late times. However, they used the linear neutrino power spectrum, neglecting the clustering arising from the nonlinear CDM potential, and were thus only able to describe the effects of neutrinos accurately in the quasi-linear regime~\cite{Viel2010}. In our simulations, neutrinos are evolved perturbatively, but clustering under the influence of the nonlinear CDM potential. Hence it is a linear response method. Other methods, such as the ones treating neutrinos as a separate particle species, a fluid component, or a ``Hybrid'' form, are discussed in detail in section~\ref{sec:sims}.

We vary three cosmological parameters: $M_\nu$, $\Omega_m$, and $A_s$. While our single resolution (512 Mpc/$h$ box size and 1024$^3$ particles) does not allow us to calibrate $P(k)$ over a wide range of scales, the relative effects of massive versus massless neutrinos are well-captured up to $k = 10 \, h$/Mpc and between redshifts $z = 0$--45. For each cosmological model, we produce a range of data products, including the simulation snapshots, halo catalogues, merger trees, galaxy weak lensing convergence maps for four source redshifts between $z_s = 1.0$--2.5, and CMB lensing convergence maps with $z_s = 1100$. Our simulations can be used as a baseline to study various effects due to massive neutrinos. 

This paper is organized as follows. We first describe the parameter sampling using the Latin Hyper-Cube scheme in section~\ref{sec:sampling}. We then describe in section~\ref{sec:sims} the $N$-body simulations, which use the public code Gadget-2 with a neutrino patch~(AB13, B17). This method allows us to compute the effect of neutrinos at relatively small additional computational cost, as compared to the large computational overhead required by particle-based neutrino simulations. We present matter power spectra in section~\ref{sec:Pmatter} and the halo catalogues and merger trees in section~\ref{sec:halo}. 
In section~\ref{sec:lens}, we describe the ray-tracing procedure used to produce lensing convergence maps for both galaxy weak lensing and CMB lensing. Finally, we summarize our conclusions in section~\ref{sec:summary}.

\section{Parameter sampling}\label{sec:sampling}

In total, we produce 100  massive neutrino  models, with three varying parameters: 

(1) $M_\nu$: the total mass of massive neutrinos,

(2) $\Omega_m=\Omega_c+\Omega_b+\Omega_\nu$: the matter density today,\footnote{We note that our definition of $\Omega_m$ includes the massive neutrinos. This definition is used in, for example, the Planck data analysis and the Boltzmann code \texttt{CAMB}. Some other analysis tools, for example the python package \texttt{astropy}, define $\Omega_m$ to be $\Omega_c+\Omega_b$ only.} and

(3) $A_s$: the primordial curvature power spectrum at the pivot scale $k_0$=0.05 Mpc$^{-1}$. \\
We {\red assume a flat universe, where $\Omega_m + \Omega_\Lambda = 1$, and} fix the Hubble parameter $h$=0.7, primordial scalar spectrum power-law index $n_s$=0.97, baryon density $\Omega_b$=0.046, and dark energy equation of state $w$=$-1$. The commonly used late-time parameter $\sigma_8$ --- the rms matter fluctuation in 8 Mpc/$h$ spheres today in linear theory --- is a derived parameter for each model. 

We assume the normal hierarchy ($m_1<m_2<m_3$), in order to reach the  minimal mass sum $M_{\nu, \rm min}\approx 0.06$~eV. For a fixed $M_\nu$, the three neutrino masses can be obtained by solving the three equations,
\footnote{Numerically, by setting $\Delta m_{21}^2\ll 1$, the equations can be solved iteratively: $\Sigma_{12} = m_1+m_2 = \frac{4}{3}M_\nu - \frac{2}{3} (M_\nu^2 + 3\Delta m + \frac{3}{4} \Delta m_{21}^2/\Sigma_{12}^2)^{\frac{1}{2}}$. The resulting mass splitting is: $m_1=0.5(\Sigma_{12} - \Delta m_{21}^2/\Sigma_{12})$, $m_2=0.5(\Sigma_{12} + \Delta m_{21}^2/\Sigma_{12})$  and $m_3 = M_\nu - \Sigma_{12}$.} 
\begin{align}
m_1 + m_2 + m_3 &= M_\nu\\
m_2^2-m_1^2 &= \Delta m_{21}^2=7.37\times10^{-5} {\rm\; eV}\\
m_3^2-\frac{m_2^2+m_1^2}{2} &= \Delta m^2=2.5\times10^{-3} {\rm\; eV\; (normal\; hierarchy}),
\end{align}
where we used the mass square differences from the 2016 Particle Data Group review~\cite{Olive:2016xmw}. 

To maximize the efficiency of sampling our multidimensional parameter space, we use the Latin hypercube method, following the Coyote~II paper~\cite{heitmann2009ii}. The goal is to spread out the points within our boundaries as evenly as possible. To do this, first imagine a cube of length 1 on each side. We then distribute 100  equally spaced points along the line connecting the [0,0,0] and [1,1,1] corners. Next, we pick 2 random points [$ x_i$, $x_j$] and 1 random coordinate $d\in$~[1,2,3], and swap their coordinate in $d$, i.e. $x_i^d\leftrightarrow x_j^d$. For the new design $\mathcal{D}$ after each swap, we evaluate the cost function,
\begin{align}
\mathcal{C}(\mathcal{D})=\frac{2D^{\frac{1}{2}}} {N(N-1)} \sum_{i=0}^{N} \sum_{j=i+1}^{N} \frac{1}{|x_i-x_j|}. 
\end{align}
where $D$ is the dimension of the parameter space and $N$ is the number of sampling points~{\red (in this work, $D=3$ and $N=100$)}. 
We repeat this process and accept the swap only if the new cost $\mathcal{C}$ is smaller than before, for 10$^4$ iterations. We use the implementation of this algorithm in the public python package \texttt{LensTools}~\cite{Petri2016Lenstools}.\footnote{\url{https://pypi.python.org/pypi/lenstools/}}

We then {\red use inverse transform sampling to} convert the uniformly sampled points into normal distributions,
centered at $\Omega_m=0.3$ and $A_s=2.1\times 10^9$, and a half-normal distribution centered at $M_\nu=0.06$~eV,\footnote{\red Note that we use $M_\nu=0.06$~eV as the center of the half-normal distribution, instead of  0.1 eV as in the fiducial model, in order to cover the minimum value for $M_\nu$.} and extend to larger values (with the prior knowledge of minimum mass sum), with standard deviations of 15\%, 15\%, and 20\%, respectively, roughly covering the estimated error size of on-going and near-future cosmological surveys, {\red such as the Dark Energy Survey, Hyper Suprime-Cam, LSST, WFIRST, Euclid, Simons Observatory, and CMB-S4.}  

\begin{figure}
\begin{center}
\includegraphics[width=1.0\textwidth]{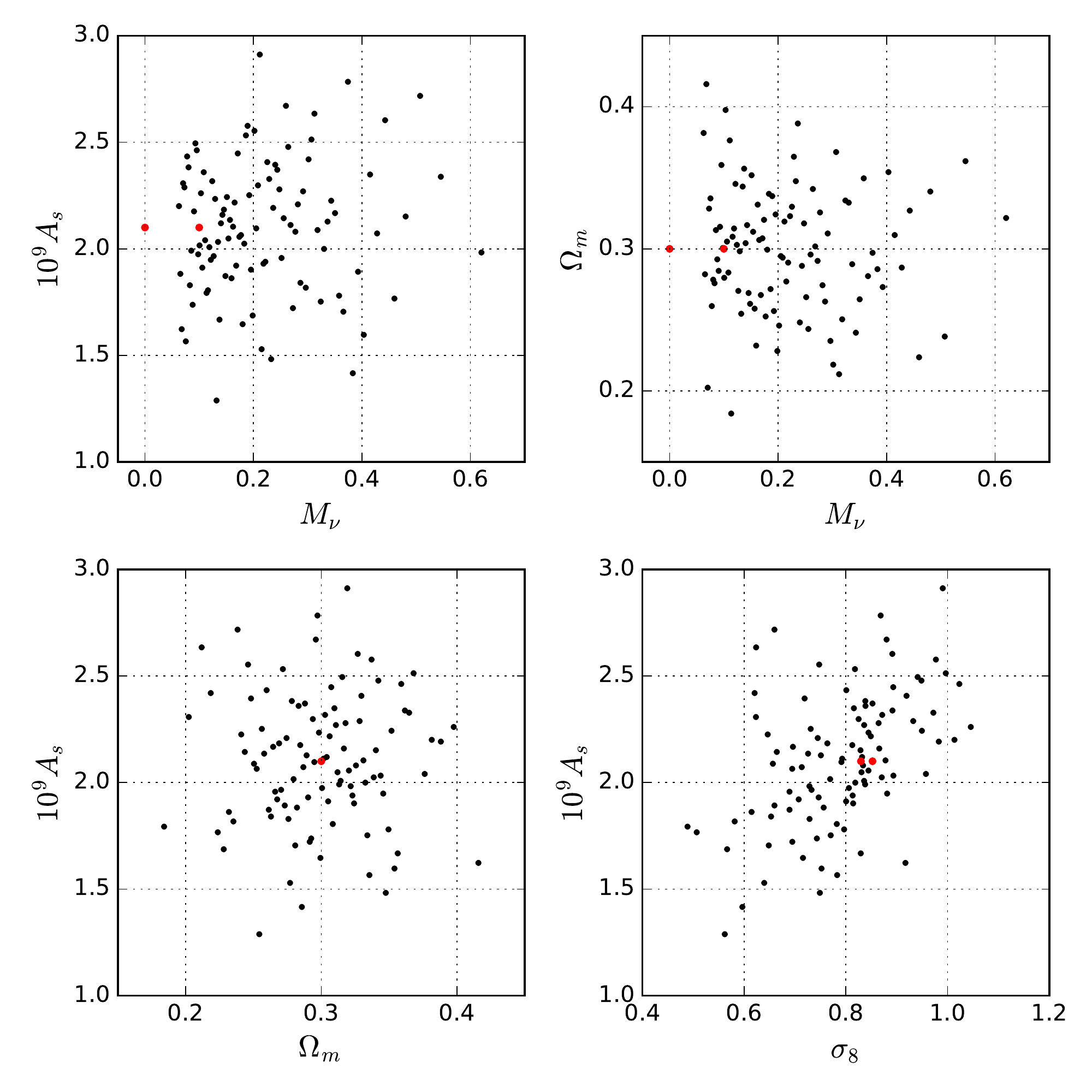}
\end{center}
\caption{\label{fig:design} The design of cosmological parameter sampling for our simulations (100 massive + 1 massless neutrino models in total). The two fiducial models ($M_\nu$=0.0~eV and 0.1~eV, $\Omega_m$=0.3, $A_s$=2.1$\times$10$^9$) are marked in red. All parameter values are listed in Table~\ref{tab: CosmoParsm}. {\red A flat universe ($\Omega_\Lambda+\Omega_m=1$) is assumed.} The other cosmological parameters are fixed at $h$=0.7, $n_s$=0.97, $\Omega_b$=0.046, and  $w$=$-1$.}
\end{figure}

We set the fiducial model to be $M_\nu=0.1$~eV, $\Omega_m=0.3$, and $A_s=2.1\times 10^9$. In addition to the 100 massive neutrino models, we also run one additional simulation with zero neutrino mass (while keeping $\Omega_m$ and $A_s$ at the fiducial model). The sampled points (100 massive models + 1 massless model) are shown in Fig.~\ref{fig:design} and listed in Table~\ref{tab: CosmoParsm}. The two fiducial models are marked in red in the figure, and  labeled as  ``1a'' and ``1b'' in the table.

\section{The simulations}\label{sec:sims}

Previously, several groups have performed simulations using a separate particle species to describe neutrinos~\cite[e.g.][]{Brandbyge2008,Villaescusa-Navarro2013,Castorina2015,Adamek2017,Emberson2017}. This method is automatically able to capture the full nonlinear neutrino clustering. However, the neutrino fluid is discretized, which causes particle shot noise, {\red with power} $P_{\rm shot} (k)=N_p^{-1/2}$, where $N_p$ is the number of particles per $k$~bin. This shot noise can swamp the true neutrino power $P_{\nu}$, especially at higher redshifts where neutrino clustering is smaller. The shot noise can only be reduced by increasing the number of neutrino particles. However, the resulting computational cost quickly becomes prohibitive\footnote{For example, the largest particle-based neutrino simulation to-date, the \texttt{TianNu} simulation (2.97 trillion particles) used more than 17 million core-hour~\cite{Emberson2017}.}. In addition, particle simulations struggle to include relativistic neutrino effects, which for low neutrino masses are relevant at early times, $z \gtrsim 25$, and cannot easily account for the neutrino hierarchy (although see \cite{Wagner2012}).

Our simulations treat neutrinos as originally laid out in AB13. The core insight is that, although the CDM density is highly nonlinear, neutrino perturbations are suppressed by their free-streaming. They can thus still be described using linear perturbation theory, provided their clustering is sourced by the full nonlinear CDM density. AB13 showed the total matter power spectrum using this approximation agreed with particle simulations at the level of 0.1\% {\red in the $k$ range covered in this work}. An advantage of this method is that neutrinos do not require a separate particle species. The computational and memory requirements for a massive neutrino simulation are thus almost identical to those for a CDM simulation. Furthermore, they do not suffer from particle shot noise. The main deficiency is that the neutrino power spectrum on small scales is not followed accurately. 
However, in this work we are interested in observable properties of the total matter and so the linear response method of AB13 is sufficient.

Further improvements in accuracy can be realized by ``hybrid'' methods, where the linear portion of the neutrinos is treated perturbatively and the nonlinear portion via particles \cite[][B17]{Brandbyge2010}. This is necessary to, e.g.~study the wake effect between neutrinos and CDM \cite{Inman2017dip}. {\red Finally, yet another} class of neutrino simulation methods are those which treat the neutrinos as a fluid with an artificial pressure, which mimics the suppression of structure caused by their peculiar velocities \cite{Inman2017,Banerjee2016}. However, the need to estimate a pressure term means that these methods are currently too slow to be useful in simulation suites such as ours.

\subsection{Initial conditions}\label{sec:ICs}

We generate our initial conditions (ICs) using a modified version of the ICs generator \texttt{N-GenIC}~\cite{springel2005}, \texttt{S-GenIC}\footnote{\url{https://github.com/sbird/S-GenIC}}, which allows us to include the radiation contribution correctly as well as massive neutrinos in Fourier space. 
It computes the factor between initial velocity and initial displacement by solving the growth equations exactly, rather than using an approximation. 

We first generate the linear matter power spectrum at $z$=99 using the Boltzmann code \texttt{CAMB}\footnote{\url{http://camb.info}}. We use the average transfer functions of CDM and baryons weighted by their respective density. A regular particle grid of collisionless CDM particles is then perturbed, where particle displacements are computed using the Zel'dovich approximation~\cite{Zeldovich1970}. 
Our ICs are at a sufficiently high redshift that the growth is linear and the effect of {\red second-order Lagrangian perturbations} is negligible. 

For all models, we use the same seed to generate the ICs, i.e. all simulations start with the same phase. This choice allows us to minimize the noise from cosmic variance when comparing two models, and also makes it possible to search for matching halos in different models.

\subsection{$N$-body simulations}\label{sec:nbody}

We use a modified version of the public tree-Particle Mesh (tree-PM)  code \texttt{Gadget-2}\footnote{\url{http://www.gadgetcode.org}}~\cite{springel2005}, with a neutrino patch. 
The background density of neutrinos includes a relativistic correction, smoothly interpolating between $a^{-3}$ and $a^{-4}$. The Fourier method to capture the massive neutrino perturbations is derived and described in detail in AB13 and B17. Here we describe briefly the simulation steps.

(1) At each time step, we evolve the particle positions with the tree-PM code. We then evaluate the particle density in Fourier space and compute the CDM power spectrum, which we use as a source term for neutrino clustering. 

(2) $P_{\nu}$ is then evaluated by integrating the displacement of a neutrino geodesic by the total matter potential, including components stored from all past timesteps (see eq.~63 in AB13). This is similar to the perturbation theory calculation done in Boltzmann codes \cite{MaBertschinger1994}, but using the nonlinear CDM matter power as a source term.

(3) The force from the neutrino over-density is added to the gravitational potential at each timestep. As we only computed the neutrino power spectrum, we assume that the Fourier phases of the neutrinos are identical to those of the CDM. Using particle simulations, B17 have confirmed that this is a good approximation.

(4) The total matter over-density $\delta_m=\rho_m/\langle\rho_m\rangle-1$, where $\langle\rho_m\rangle$ is the mean matter density, is then evaluated as
\begin{align}\label{eq:delta_m}
\delta_m({\bf k}, t) &= (1-f_\nu)\delta_{\rm \red c+b}({\bf k},t)+f_\nu\delta_\nu({\bf k},t)  {\rm , \; where}\\
f_\nu&=\frac{\Omega_\nu}{\Omega_m}=\frac{1}{\Omega_m h^2}\frac{M_\nu}{93.14~{\rm eV}}
\end{align}

Our resolution is 1024$^3$ particles with box size 512~Mpc/$h$, corresponding to a mass resolution of $\approx10^{10} M_\odot/h$.  We run one box per model, with periodic boundary conditions. Snapshots are output continuously every 180~Mpc (or 126~Mpc/$h$), from $z$=45 to $z$=0.
The choice of the highest redshift $z$=45 is determined by requiring that the redshift range covers 99\% of the structure growth weighted by the CMB lensing kernel, where we use the linear growth factor $D(z)\sim a$. We do not include any relativistic corrections in our method, which only make sub-percent differences on very large scales ($k < 0.005 \, h$/Mpc)~\cite{Adamek2017},
and are hence irrelevant to our simulations with $k_{\rm min}\approx 0.01 \, h$/Mpc.

\section{Matter power spectrum}\label{sec:Pmatter}

\begin{figure}
\begin{center}
\includegraphics[width=1.0\textwidth]{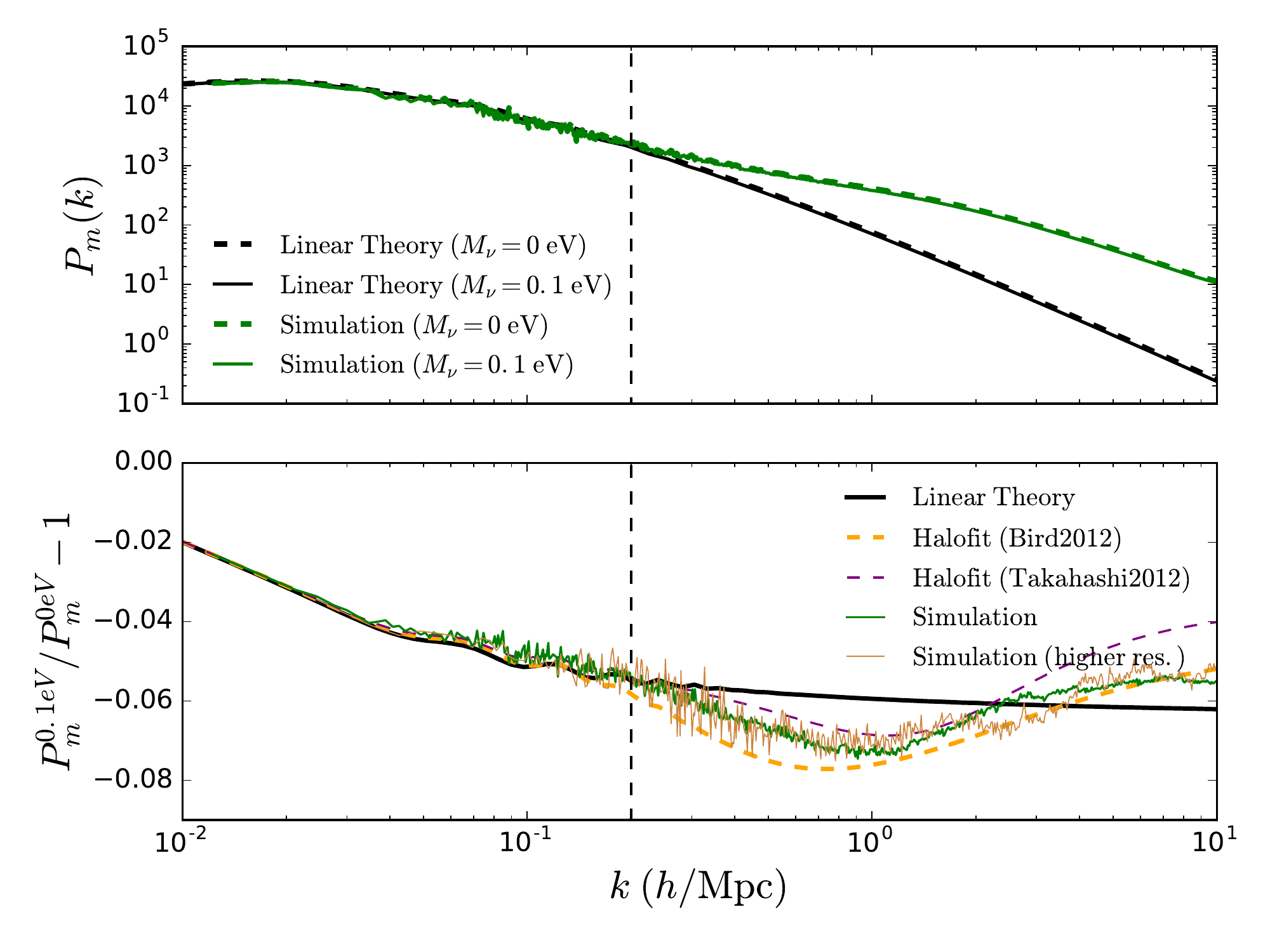}
\end{center}
\caption{\label{fig:pmatter_fidu} {\bf Upper}: the matter power spectra for the fiducial models, where $\Omega_m$=0.3, $A_s$=2.1$\times10^9$, and $M_\nu=0.1$~eV (massive, solid curves) and 0 eV (massless, dashed curves). Other cosmological parameters are fixed at $h$=0.7, $n_s$=0.97, $\Omega_b$=0.046 and  $w$=$-1$. The simulations (green curves) have 1024$^3$ DM particles and box size 512~$h$/Mpc. We also show the theory curves from linear theory for comparison (black curves). {\bf Lower}: the fractional difference between the $P_m$ of the massive model and that of the massless model, {\red i.e.}, the suppression due to massive neutrinos, measured in our simulations (green solid curves) and compared with linear theory (black solid curves) and two versions of \texttt{Halofit} (dashed curves). We also show additional higher-resolution runs, with the same number of particles but half the box size (labeled as ``higher res.'', brown solid curve). The dashed vertical lines denote the approximate division between linear and nonlinear scales.}
\end{figure}

We show the matter power spectra  $P_{m}(k)$ of the two fiducial models at $z=0$ in Fig.~\ref{fig:pmatter_fidu}. In the upper panel, we compare the simulations with linear theory, and find excellent agreement on linear scales ($k\le 0.2 \, h$/Mpc). The difference on small scales is due to  nonlinear growth. In the lower panel, we show the fractional difference between the massive and massless fiducial models for the simulations, linear theory, and two versions of \texttt{Halofit}~(\cite{Bird2012} and~\cite{takahashi2012}, both based on the original version from~\cite{smith2003}). Again, we see good agreement of all curves on large scales. On small scales, the nonlinear suppression due to massive neutrinos is clearly seen (for example, an extra 1--2\% fractional suppression at $k = 1 \, h$/Mpc compared with linear theory).  The discrepancy between the \texttt{Halofit} models cautions us that these models have  limited accuracy at present.

As a convergence test, we also run two additional simulations for the massive and massless fiducial models, but with a higher resolution of 1024$^3$~particles and a 256~Mpc/$h$ box size (the same number of particles, but half the box size of our fiducial runs). We show the matter power spectrum difference measured from these two runs, labeled as ``higher res.'', in the lower panel of Fig.~\ref{fig:pmatter_fidu} (their $P_m$ are not shown in the upper panel, as they are indistinguishable by eye from the fiducial runs). We see good agreement between our fiducial runs and the higher resolution runs for $k$=0.01--10~$h$/Mpc. 

\begin{figure}
\begin{center}
\includegraphics[width=1.0\textwidth]{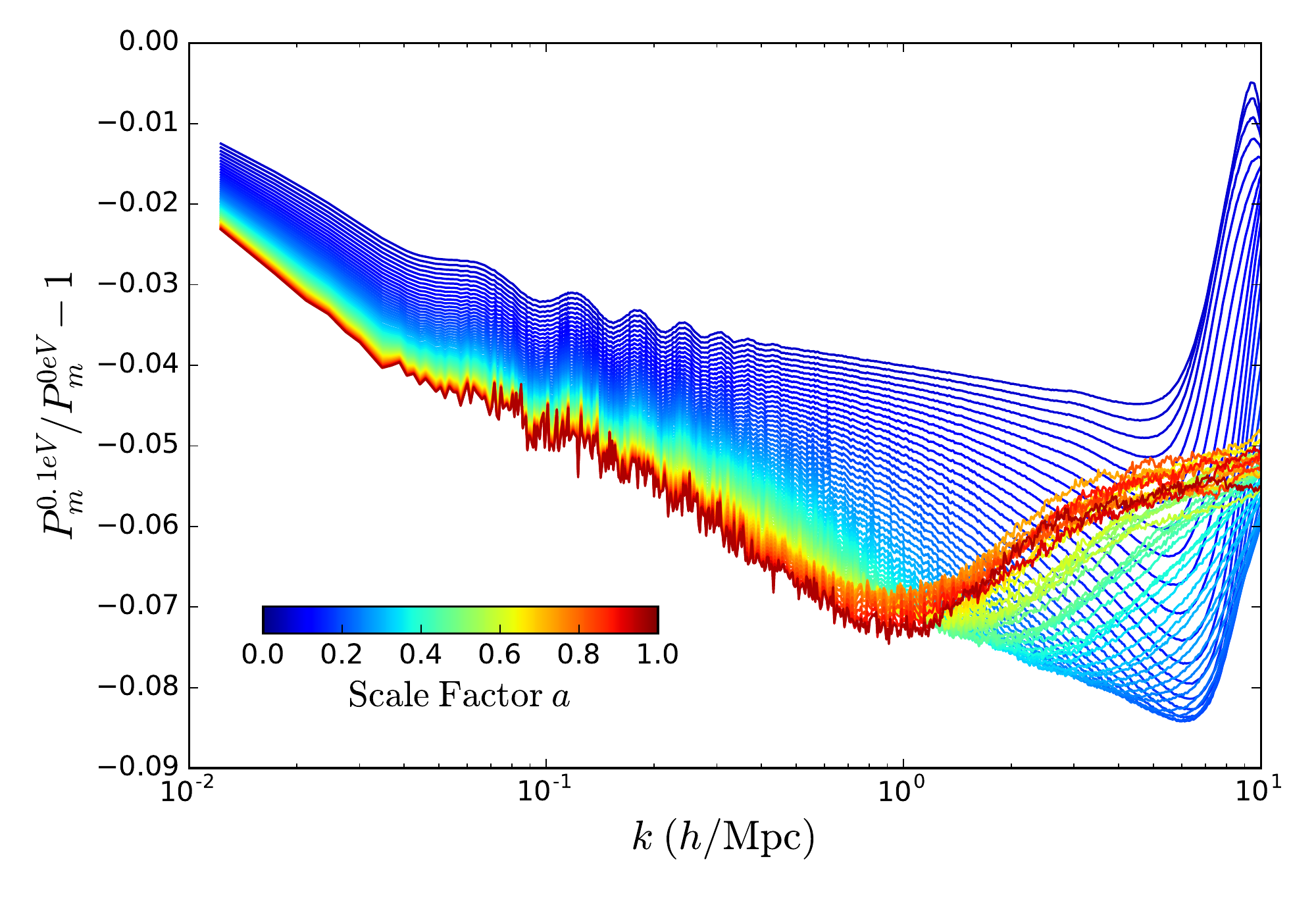}
\end{center}
\caption{\label{fig:pmatter_evolve} Evolution of the suppression of the matter power spectrum due to massive neutrinos of 0.1~eV, measured in our simulations. The ``spoon'' shape seen in the nonlinear regime is directly linked to the change in the halo mass function.}
\end{figure}

With continuous redshift outputs from our simulations, we can also see the interesting redshift evolution of the power suppression ($P_{m}^{0.1eV} / P_{m}^{0 eV} - 1$), shown in Fig.~\ref{fig:pmatter_evolve}. On linear scales, the suppression becomes larger as the redshift decreases. On nonlinear scales, we see the shift of the minimum of the ``spoon'' shape~\cite{Brandbyge2008,Viel2010,Agarwal2011,Bird2012,Wagner2012,Massara2014} from around $k = 6 \, h$/Mpc at $z>10$ to $k=1 \, h$/Mpc at $z=0$. The spoon shape is not captured in linear theory, but can be understood using the halo model~\cite{cooray2002}. In this $k$ range, the matter power spectrum receives contributions mostly from the 1-halo term, which describes the clustering of matter within a single halo. The $k$-range that a halo can impact depends on its size, and hence its mass, because it can not contribute to scales larger than its size. The upturn at high $k$ means the small halos are less impacted by massive neutrinos, which is also apparent in Fig.~\ref{fig:hmf} (see discussion in the next section on the halo mass function). The minimum of the spoon corresponds to the size of the most massive halos, which are less abundant in the presence of massive neutrinos. The shift of the minimum to smaller $k$ towards $z=0$ is due to the hierarchical structure formation, where larger halos contributing to this $k$ range form only at low redshifts. 

In addition, we show the difference between all massive models from the massless fiducial model in Fig.~
\ref{fig:pmatter_all}. We compare the measured $P_{m}$ from simulations (solid curves) to that from linear theory (dashed curves), and find excellent agreement for all models on large scales. The actual shape for each model depends on the interplay of all three input parameters (shown in the labels).

\begin{figure}
\begin{center}
\includegraphics[width=0.92\textwidth]{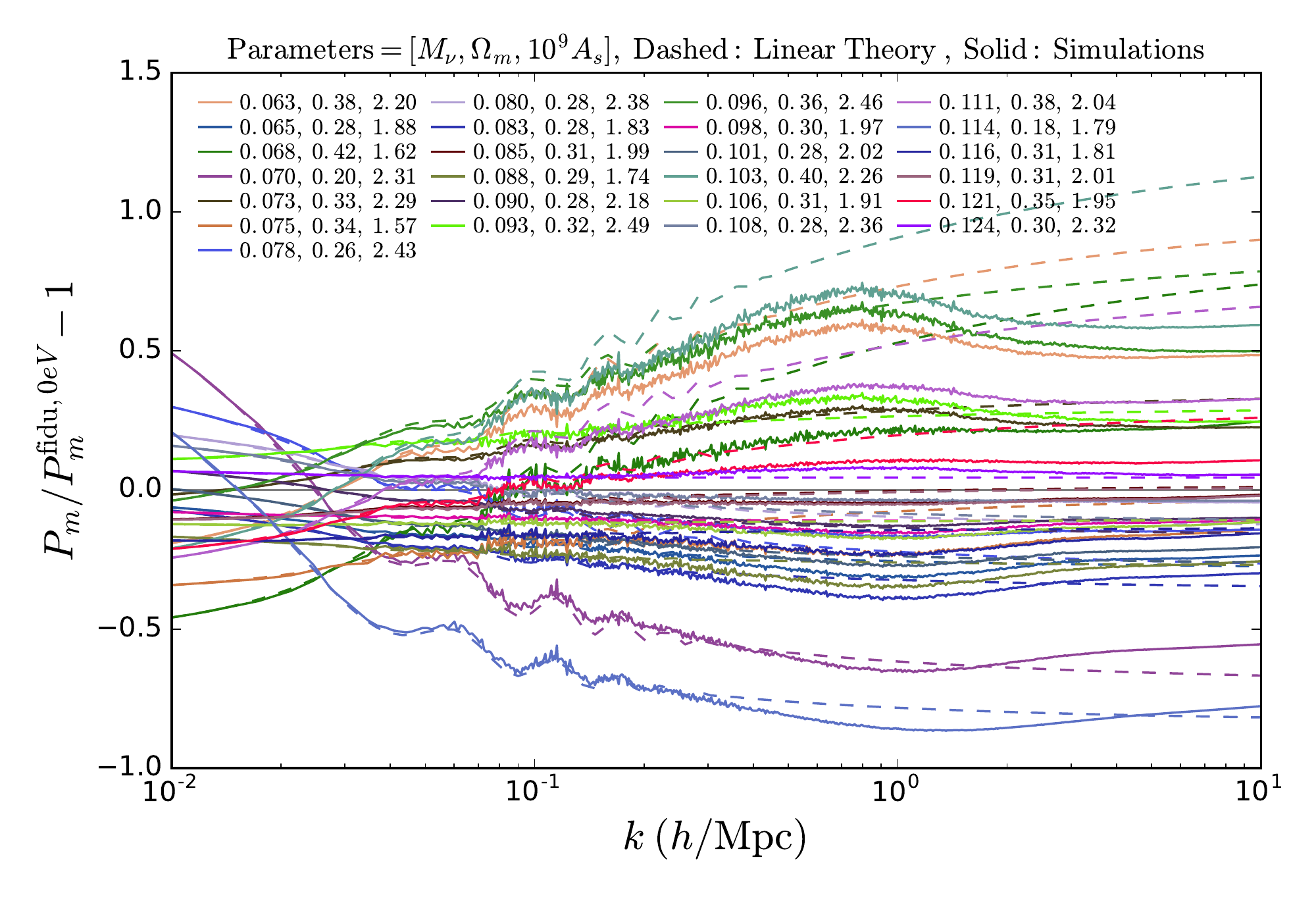}
\includegraphics[width=0.92\textwidth]{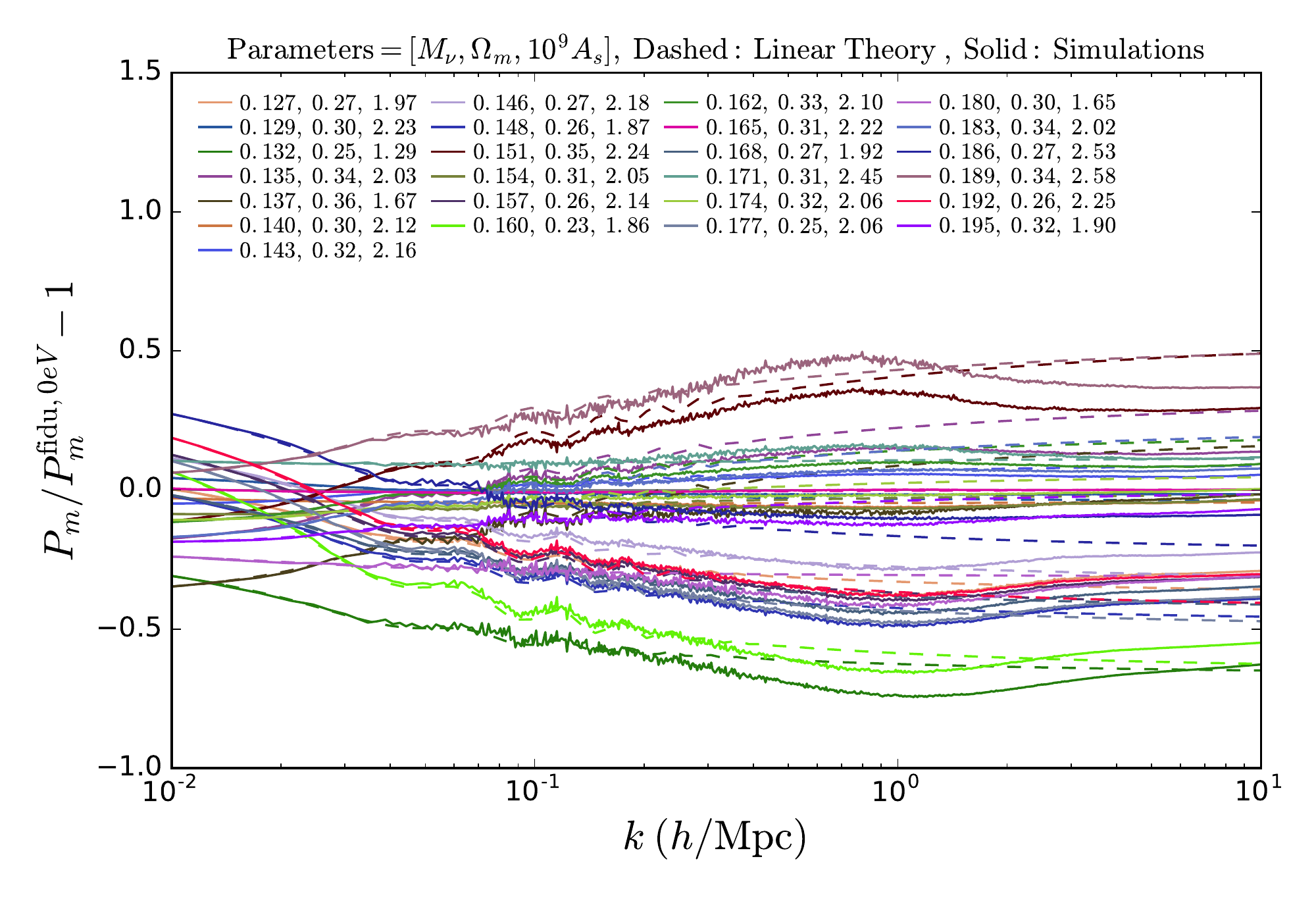}
\end{center}
\caption{\label{fig:pmatter_all} The difference of the matter power spectra of massive models from the massless fiducial model ($z$=0), {\red shown for both simulations (solid curves) and linear theory (dashed curves)}. Parameters are listed in Table~\ref{tab: CosmoParsm}.}
\end{figure}

\begin{figure}\ContinuedFloat
\begin{center}
\includegraphics[width=0.92\textwidth]{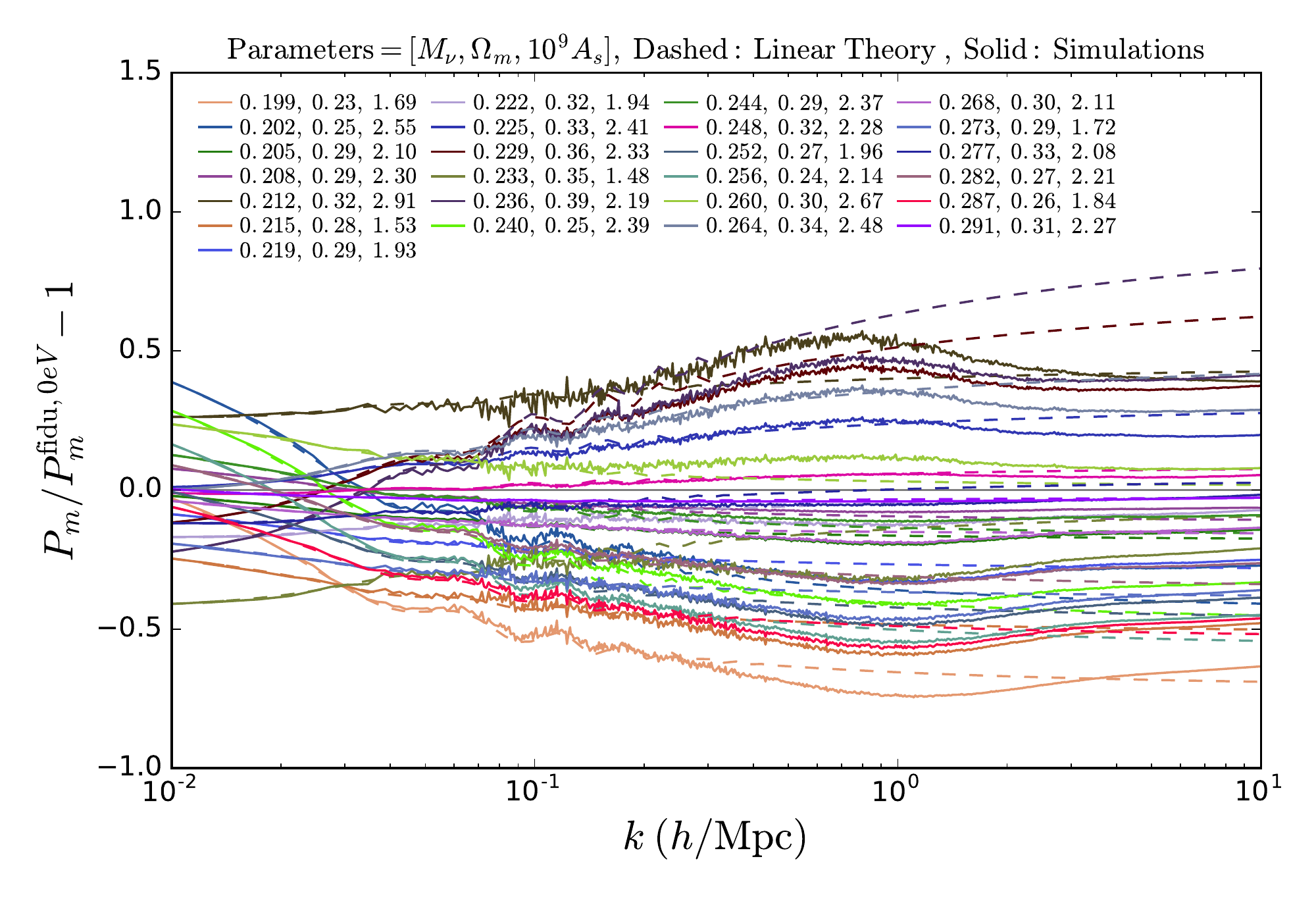}
\includegraphics[width=0.92\textwidth]{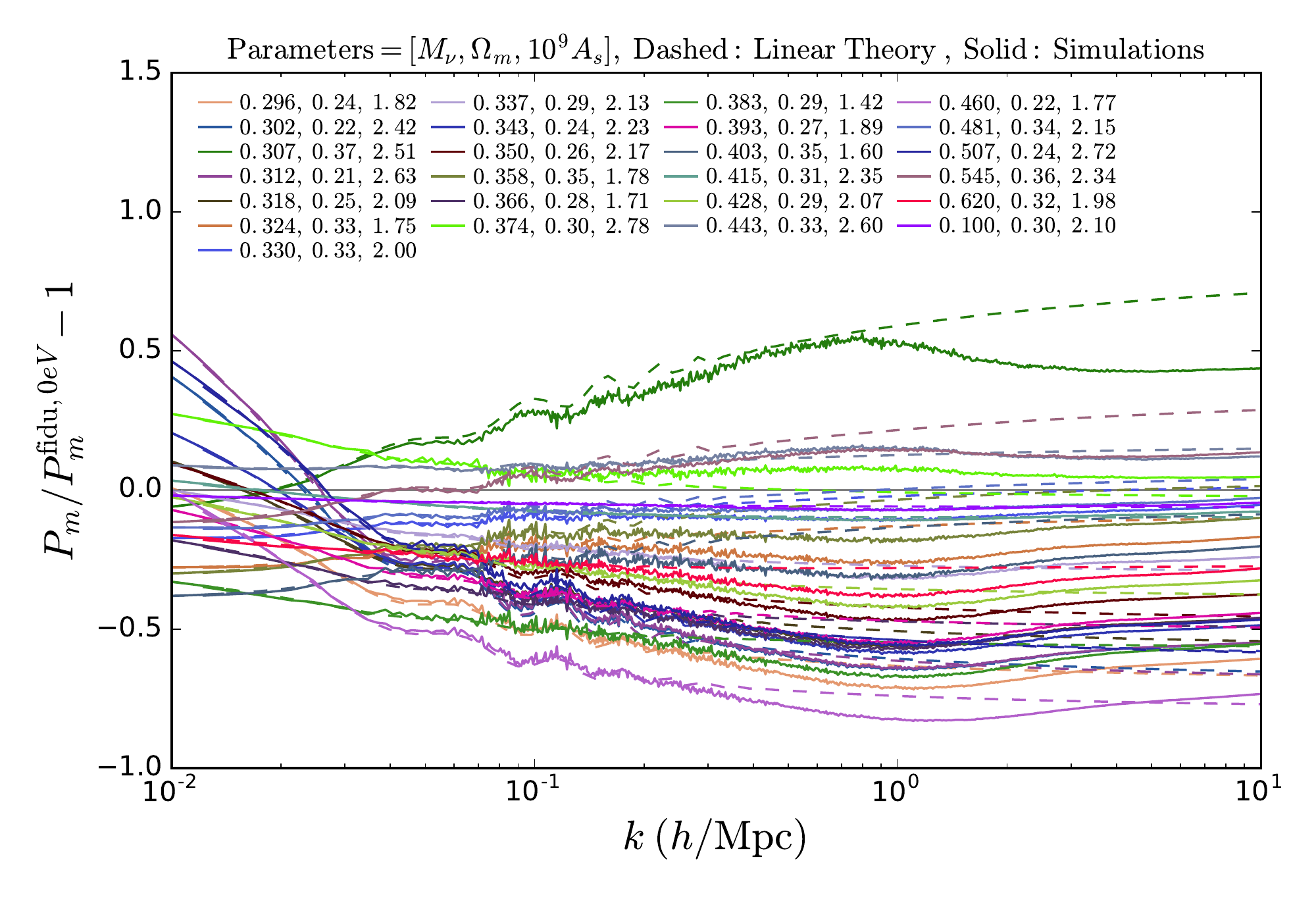}
\end{center}
\caption{\label{fig:pmatter_all}   (Cont.) The difference of the matter power spectra of massive models from the massless fiducial model ($z$=0), {\red shown for both simulations (solid curves) and linear theory (dashed curves)}. Parameters are listed in Table~\ref{tab: CosmoParsm}.}
\end{figure}

\section{Halo catalogue and merger trees}\label{sec:halo}

We generate halo catalogues for each snapshot for all 101 cosmological models, using the public halo finder code \texttt{Rockstar}\footnote{\url{https://bitbucket.org/gfcstanford/rockstar}}~\cite{Behroozi2013Rockstar}. \texttt{Rockstar} uses a friends-of-friends-based algorithm, where particles physically nearby  
are first identified and grouped. Substructures are then searched for within the parent halos. The code has the advantage of using all six phase-space (position and velocity) dimensions and one temporal dimension. Our boxes are output densely in the redshift dimension ($\approx70$ redshifts per model from $z=45$ to $z=0$), hence the inclusion of the temporal information in the halo finder ensures the consistency of halo properties across all time steps. It also reduces the parent halo--subhalo ambiguity in major mergers normally faced by halo finders that only operate on single snapshots.

We record the standard halo properties, including, for example, {\red the halo mass $M_{\rm vir}$, radius $r_{\rm vir}$, maximum circular velocity, velocity dispersion, position, velocity, spin~\cite{Peebles1969}, and shape (ratios of the principal axes)}. Descriptions are included in the header of each halo catalogue file. The minimal halo mass recovered in our simulations is $\approx 10^{11}M_\odot$. We note that, because the massive neutrinos are computed analytically and because the halo finder can only search for CDM particles, the halo mass in our catalogues omits the contribution from neutrinos. We expect this to have a negligible effect because only a relatively small fraction of neutrinos are bound to massive halos, while a majority remain weakly clustered (see discussion in AB13).


\begin{figure}
\begin{center}
\includegraphics[width=1.0\textwidth]{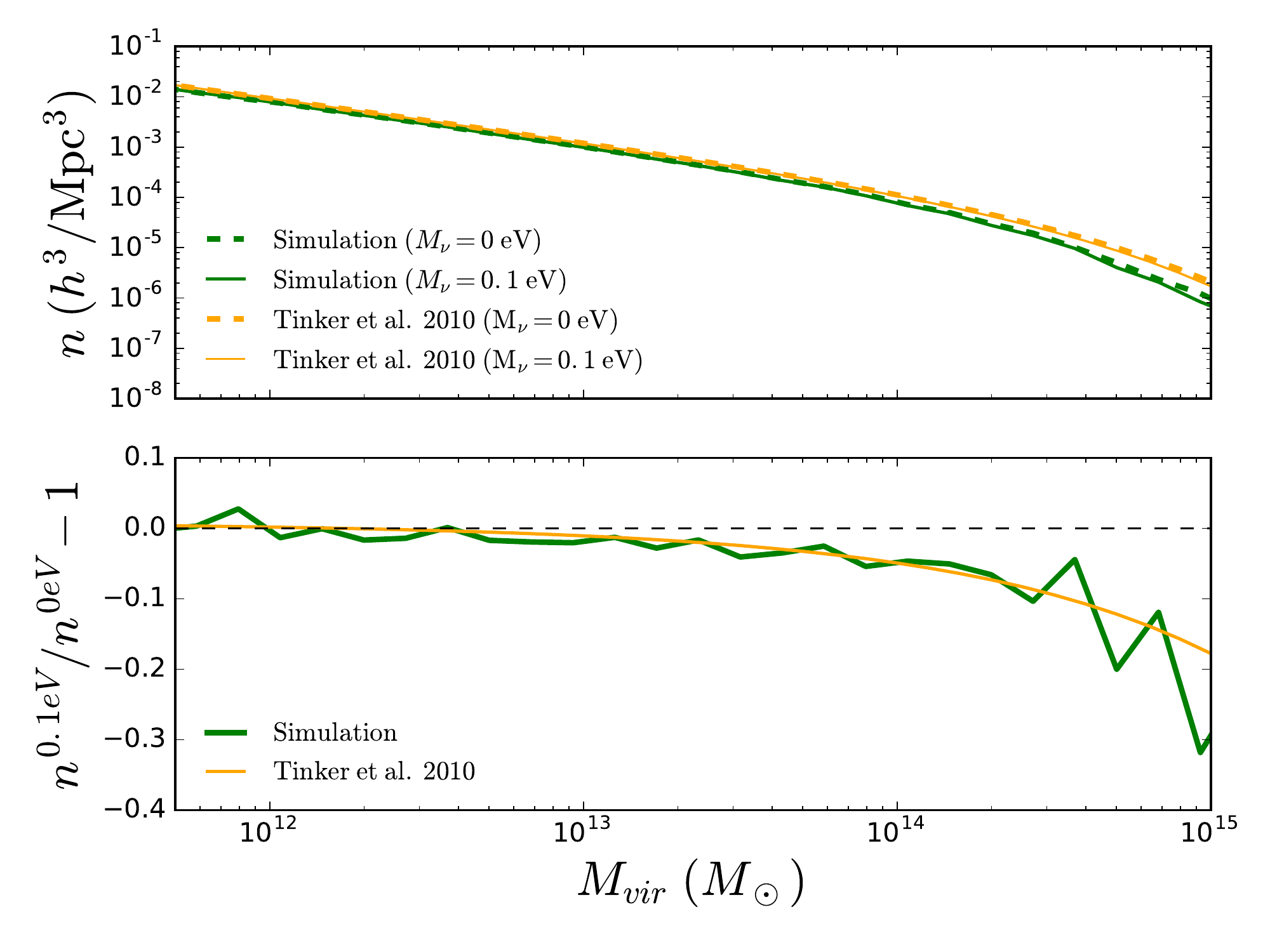}
\end{center}
\caption{\label{fig:hmf} {\bf Upper}: the halo mass function for the the fiducial models, where $\Omega_m = 0.3$, $A_s = 2.1 \times 10^9$, and $M_\nu = 0.1$~eV (massive) and 0 eV (massless). The other cosmological parameters are fixed at $h=0.7$, $n_s=0.97$, $\Omega_b=0.046$, and  $w=-1$. We also show the fitting function from Tinker et al. (2010)~\cite{Tinker2010}. {\bf Lower}: fractional difference between the halo mass function in the massive neutrino model and the massless model.}
\end{figure}

We show the halo mass functions from our fiducial simulations for the $z=0$ snapshot in Fig.~\ref{fig:hmf}, compared with the fitting formula from Tinker et al. (2010)~\cite{Tinker2010}, where we use the nonlinear matter power spectrum from the \texttt{Halofit} model~\cite{smith2003,takahashi2012} to calculate the matter variance~(see Eq.~3 in~\cite{Tinker2010}).
The number density of halos of mass $>10^{14} M_\odot$ is lower in our simulation than in the fitting formula, likely due to our limited box size. In the lower panel, we show the fractional difference between the halo mass function of a massive neutrino model and that of the massless neutrino model, and find excellent agreement between our simulations and the fitting formula. While low-mass halos are less impacted by the massive neutrinos (0.1~eV), the number of high-mass ($>$ a few $\times 10^{14} M_\odot$) halos is reduced by 10--20\%. Because massive neutrinos smooth the small-scale density field, the number of rare high-density peaks is suppressed as a result.  This directly causes the spoon shape and its evolution seen in Fig.~\ref{fig:pmatter_evolve}. This effect is similar to a massless neutrino model that has a smaller value of $\sigma_8$, {\red although the effects are not exactly degenerate, as discussed in \cite{Ichiki2012}}.

Next, we create merger trees using the public code \texttt{Consistent Tree}\footnote{\url{https://bitbucket.org/pbehroozi/consistent-trees}}~\cite{Behroozi2013Tree}, a companion code to \texttt{Rockstar}. Similarly, it takes advantage of temporal information to secure the consistency of halo properties across multiple time steps. \texttt{Consistent Tree} first takes the traditional method of finding common particles to match progenitors and descendants in consecutive snapshots. It then makes the assumption that each descendant must have a progenitor (except for when the progenitor mass falls below the mass resolution), and traces back in time to find the best-matching progenitor by evolving gravity backwards. When progenitors are missing in an intermediate time step, often due to their close passage near the center of its host halo (therefore mis-identified as a part of the host halo), \texttt{Consistent Tree} creates the halo and assigns properties interpolated using the information from adjacent time steps.

\begin{figure}
\begin{center}
\includegraphics[width=1.0\textwidth]{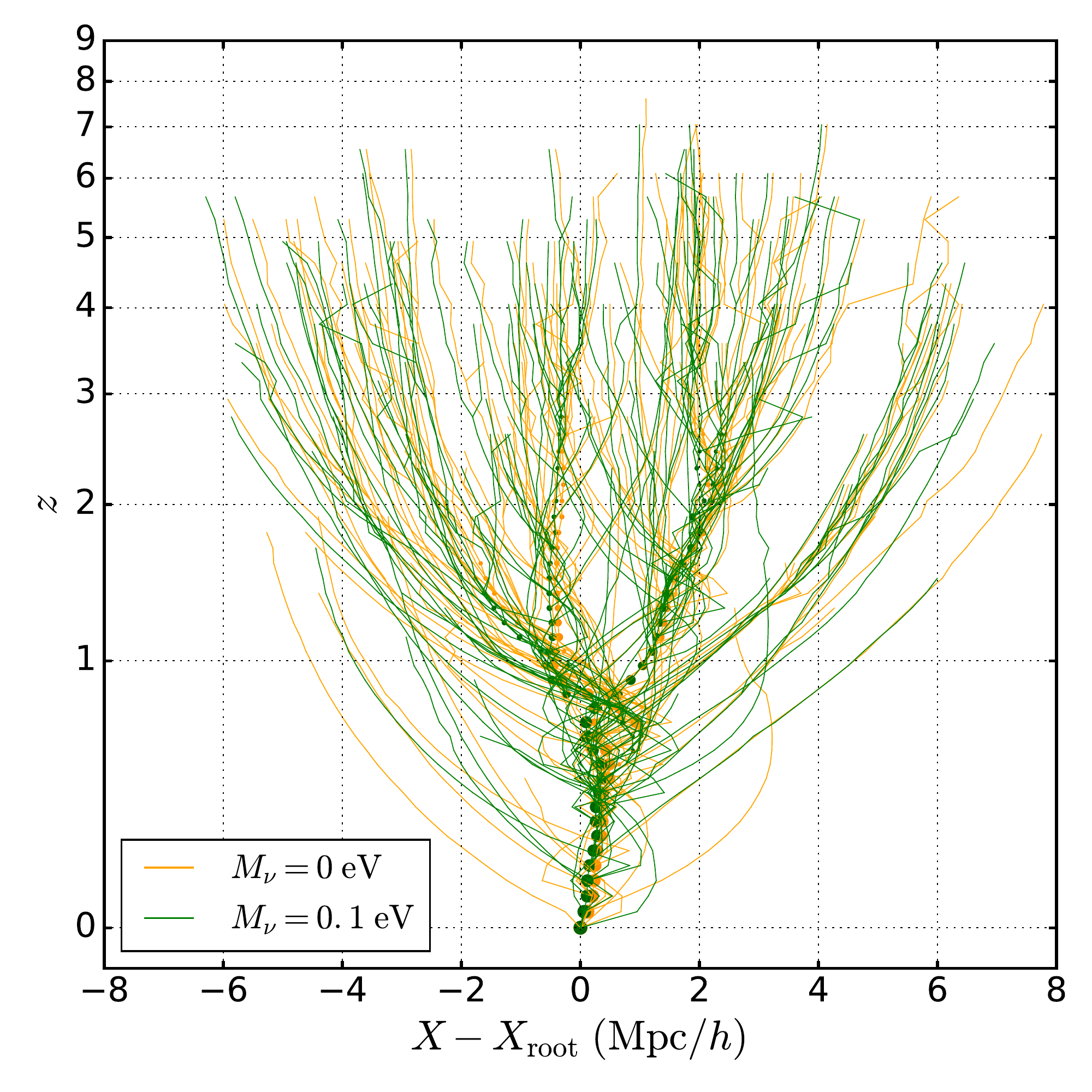}
\end{center}
\caption{\label{fig:tree} Example of the merger tree for a $M_{vir}=10^{14.5}M_\odot$ halo, in the massive (green) and massless (orange) fiducial simulations. $X$ is the projected halo position on the (comoving) x-axis, and $X_{\rm root}$ is the final position of the halo in the $z$=0 snapshot. Massive halos ($>10^{12.5}M_\odot$) are marked by solid circles, with size corresponding to the halo mass. }
\end{figure}

We show an example of the merger history of a $M_{vir}=10^{14.5}M_\odot$ halo in Fig.~\ref{fig:tree}, in both the massive and massless fiducial simulations. Because we use the same seed to generate the initial conditions for all models, we can match the same halos across simulations based on their mass and position. This is more reliable for very massive halos, since small halos may be destroyed in some models. We see many clearly matching branches in both simulations, especially for all the major branches where the most massive progenitors ($>10^{12.5}M_\odot$, solid circles) are formed. However, some branches may only exist in one of the models; for example, the outer-left-most branch in the massless simulation.  

\section{Lensing convergence maps}\label{sec:lens}

Weak gravitational lensing by large-scale structure is a promising cosmological probe. Compared to other cosmological probes, which typically observe the baryon distribution and then infer the underlying DM distribution, weak lensing has the advantage of probing the DM clustering directly. Photons emitted at cosmological distances are deflected by the intervening matter. As a result, we see a distorted image of the source.  Lensed galaxies are magnified in brightness and distorted~(``sheared'') from their intrinsic shape~(see a recent review by~\cite{Kilbinger2015}). Lensing distortions produce non-Gaussianity in CMB maps of temperature and polarization anisotropies~(see a recent review by~\cite{Hanson2010}). Statistical measurements of CMB lensing~\cite[e.g.][]{Das2011,vanEngelen2012,Planck2015XV,Sherwin2016ACTPol,Omori2017} and galaxy weak lensing~\cite[e.g.][]{Schrabback2010,Heymans2012,Hildebrandt2017,Mandelbaum2017,2017DES} have been achieved recently, and are now advancing to become a major tool for precision cosmology.

To lowest order, the lensing convergence $\kappa$ is a projection of the three-dimensional matter over-density $\delta_m$ along the line-of-sight, weighted by the lensing kernel $W(z)$, which describes the efficiency of lenses at each redshift in deflecting the source light. Under the Born approximation~\cite{Hilbert2009,Schafer2012,Pratten2016,Petri2017}, where photons are assumed to travel along unperturbed geodesics $\xB=\chi(z)\thetaB$, we can write,
\begin{align}
\label{eq:kappadef}
\kappa(\thetaB) &= \int_0^{\infty} dz W(z) \delta(\chi(z)\thetaB, z), {\rm \; where}\\
\label{eq:kernel}
W(z) &= \frac{3}{2}\Omega_{m} H_0^2 \frac{(1+z)}{H(z)} \frac{\chi(z)}{c} \times\int_z^{\infty} dz_s \frac{dn(z_s)}{dz_s} \frac{\chi(z_s) - \chi(z)}{\chi(z_s)}
\end{align}
where $\chi(z)$ is the comoving distance, $\thetaB$ is the angular position, $z_s$ is the source redshift, and $dn(z_s)/dz_s$ is the redshift distribution of the sources and is a delta function at $z_s=1100$ for the CMB. {\red Ref.~\cite{Petri2017} showed that the Born approximation is sufficient to describe the lensing convergence power spectrum, but breaks down at higher orders. Therefore, ray-tracing simulations, such as done in this work, are necessary to capture the full statistics of the lensing field.}

To generate lensing convergence maps, we first slice the simulation boxes to create density planes, and then ray-trace through the planes from $z=0$ to the source redshifts. We use the public code \texttt{LensTools}~\cite{Petri2016Lenstools} for both lens plane construction and ray-tracing. We describe in detail our procedure below.

\subsection{Density planes and ray-tracing}

For each snapshot, we cut the box into 4 planes, each with  (comoving) thickness 180~Mpc, equal to the output interval between snapshots (or 126~Mpc/$h$, compared to the box size of 512~Mpc/$h$).  Applying this to all three dimensions, we are able to generate 12 planes per snapshot. Particles in each plane are projected on to a two-dimensional CDM density plane. To get the total matter density, we add the neutrino component following Eq.~\ref{eq:delta_m}, using the $P_\nu$ computed from the simulation and assuming the same Fourier phases for neutrinos and CDM particles. The gravitational potential is then calculated using the Poisson equation. 

\begin{figure}
\begin{center}
\includegraphics[width=1.0\textwidth]{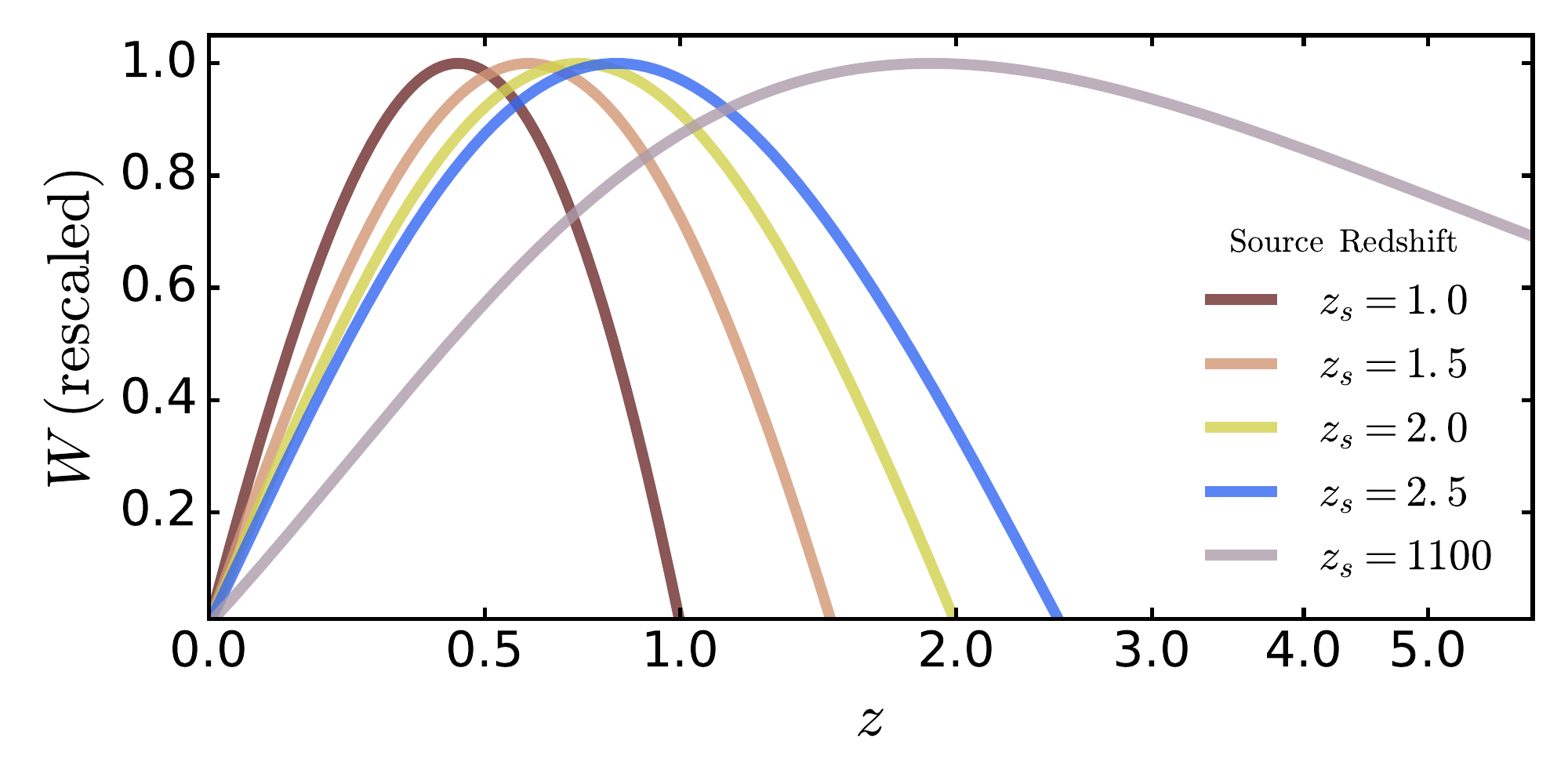}
\end{center}
\caption{\label{fig:kernel} Lensing kernels (eq.~\ref{eq:kernel}) for all five source planes, located at $z_s=1.0, 1.5, 2.0, 2.5$, and 1100. They describe the sources' sensitivity to the intervening matter fluctuations as a function of redshift. The curves are rescaled for clarity.}
\end{figure}

Next, we shoot $4096^2$ {\red regularly spaced} light rays from the center of the $z=0$ plane backwards in time, spreading over a 3.5$\times$3.5 deg$^2$ solid angle, and track the trajectories of each light ray until the source plane. Light rays travel in straight lines between planes, and are deflected at each plane. The deflection angle and convergence are calculated at each plane, following~\cite{Jain2000,Hilbert2009} (see Fig.~1 of \cite{Hilbert2009} for a clear illustration). During this procedure, it is never assumed that the deflection angle is small or that the light rays follow unperturbed geodesics. Therefore, our convergence maps can be used to test the so-called ``post-Born'' corrections~\cite{Schafer2012,Pratten2016,Petri2017}. We ray-trace to five different source planes: $z_s =1.0, 1.5, 2.0, 2.5$ for galaxy lensing, and $z_s=1100$ for CMB lensing. The lensing kernels (eq.~\ref{eq:kernel}) for all our source planes are shown in Fig.~\ref{fig:kernel}, and are rescaled for clarity. 

\begin{figure}
\begin{center}
\includegraphics[width=1.0\textwidth]{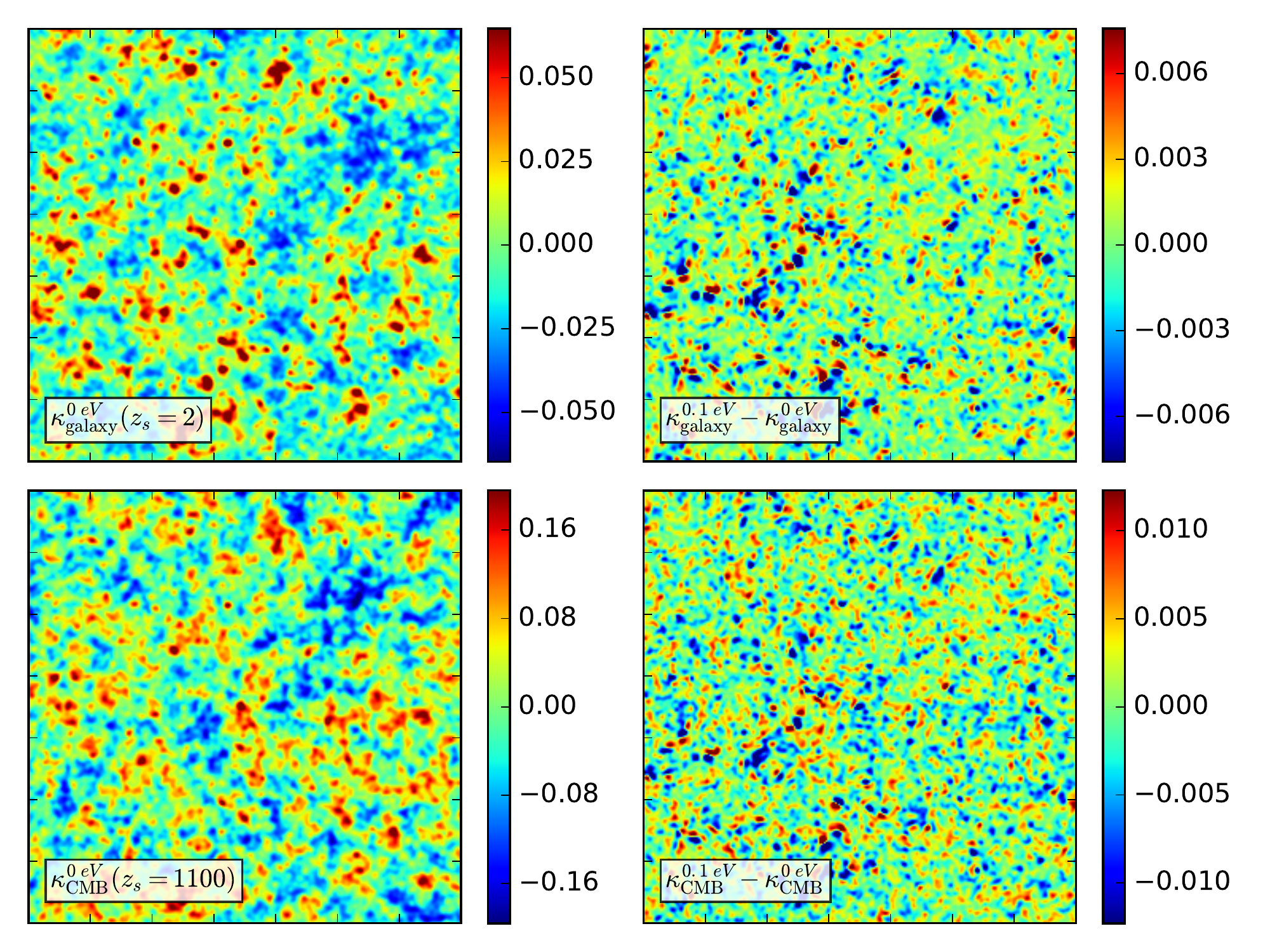}
\end{center}
\caption{\label{fig:conv} Sample $\kappa$ (lensing convergence) maps from the fiducial massless-neutrino model for galaxy sources at $z_s=2$ (upper left) and the CMB (lower left), and corresponding differences between the massive and massless models (right panels). Similar features are seen in both $\kappa$ maps due to their overlapping lensing kernels. Correlations can also be seen between the difference maps and their corresponding $\kappa$ maps at left.}
\end{figure}

\begin{figure}
\begin{center}
\includegraphics[width=1.0\textwidth]{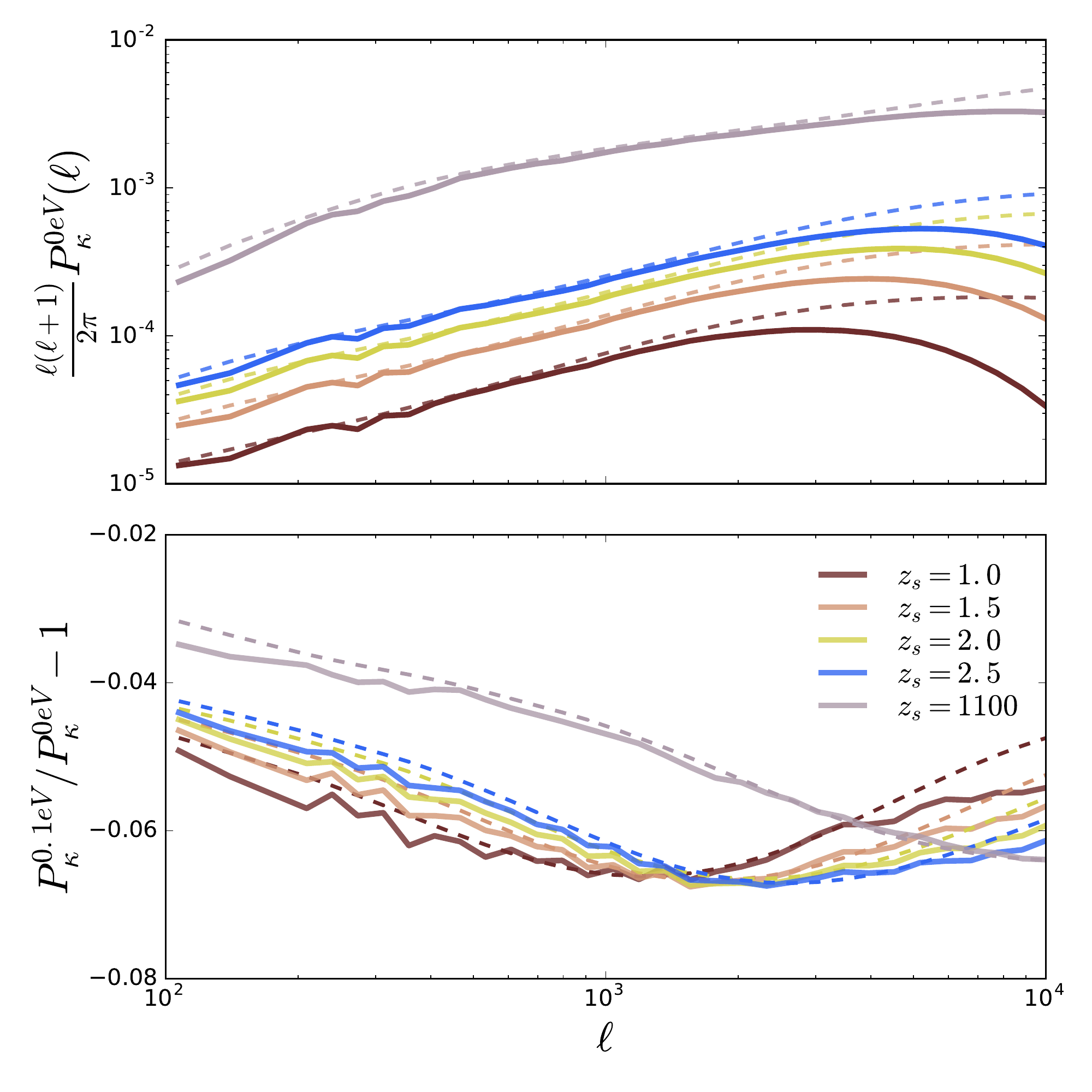}
\end{center}
\caption{\label{fig:pkappa} {\bf Upper}: Convergence power spectra ($P_{\kappa}$) for the fiducial massless-neutrino model using five source redshifts measured from our simulations (solid curves), compared with the \texttt{Halofit} fitting formula from~\cite{smith2003} and~\cite{takahashi2012} (dashed curves). The massive-neutrino fiducial model power spectra are not shown for clarity, as they are very similar to the massless ones. {\bf Lower}: The fractional difference of the massive-model $P_{\kappa}$ from the massless-model $P_{\kappa}$, for each source redshift, for both the simulations (solid curves) and the \texttt{Halofit} model (dashed curves). Each simulation $P_\kappa$ is averaged over 1000 realizations.}
\end{figure}

\subsection{Galaxy and CMB lensing convergence maps}

We generate 1,000 convergence ($\kappa$) map realizations per source redshift, for each cosmology, by randomly rotating and shifting the potential planes. {\red Ref.~\cite{Petri2016a} has demonstrated that one $N$-body simulation is sufficient to generate few$\times$$10^4$ statistically independent realizations.} The $\kappa$ maps are 2048$^2$ pixels and 3.5$^2$=12.5 deg$^2$ in size, with square pixels of side length 0.1025 arcmin. We do not include observational noise in our simulated maps, from for example galaxy shape noise or CMB detector noise. The noise contributions should be modeled for specific survey designs. For example, galaxy shape noise can be assumed to follow a Gaussian distribution, with a width related to the telescope's {\red point spread function} and observed galaxy number density. CMB lensing reconstruction noise from the primordial CMB, detectors, atmosphere, and foregrounds can be included using public tools such as \texttt{Quicklens}\footnote{\url{https://github.com/dhanson/quicklens}}.

We show one $\kappa$ map realization in the massless-neutrino fiducial model, for source redshifts $z_s=2$ and $z_s=1100$ in the upper- and lower-left panels of Fig.~\ref{fig:conv}, respectively. We do not show the corresponding $\kappa$ maps for the massive-neutrino fiducial model as they look almost identical to the massless maps. Instead, we show the difference between the $\kappa$ maps for the massive and massless models in the right panels. Similar features are seen in both the galaxy and CMB $\kappa$ maps, due to the overlapping kernels of both maps. The difference maps in the right panels also show {\red hints of} correlations with their corresponding $\kappa$ maps. 
{\red For example, there appears to be a mild excess in void regions.}  

We show the convergence power spectra $P_{\kappa}(\ell)$ for the massless fiducial model measured from our simulations (solid curves) and that predicted by the \texttt{Halofit} model~\cite{smith2003,takahashi2012} (dashed curves) in the upper panel of Fig.~\ref{fig:pkappa}. The deficit of power in the simulation at high $\ell$ is due to our limited resolution, and is worse at lower source redshifts, as the same physical object extends over a larger angular scale when closer.  There is also a small deficit at low $\ell$ due to the finite box size. The suppression in  $P_{\kappa}(\ell)$ due to massive 0.1~eV neutrinos is shown in the lower panel, again for both simulations (solid curves) and the \texttt{Halofit} model (dashed curves). We see a general agreement with \texttt{Halofit}, with small discrepancies, likely due to the intrinsic difference between our matter power spectra as seen in Fig.~\ref{fig:pmatter_fidu}.

While our $\kappa$ maps are relatively small in size (12.25 deg$^2$), they can be very useful to study structure growth in the nonlinear (or non-Gaussian) regime, which is more prominent on small scales. Previous analyses of non-Gaussian statistics for galaxy weak lensing surveys have shown that there is comparable information in higher-order statistics, such as peak counts, Minkowski Functionals, and higher-order moments, compared to that in the power spectrum~\cite[e.g.][]{shirasakiyoshida2014,Liu2015,Petri2015,Liux2015,Kacprzak2016,Martinet2017,Shan2017}. Forecasts of non-Gaussian statistics for CMB lensing have also shown useful constraining power when combined with the power spectrum~\cite[e.g.][]{Namikawa2016,Liu2016b}, although smaller than for galaxy weak lensing surveys due to the higher source redshift in CMB lensing.

In addition, as we use the same seed to generate each realization for all source redshifts, our $\kappa$ maps for the five source redshifts~(see Fig.~\ref{fig:kernel}) are correlated. This is particularly useful for studying lensing tomography, as well as the cross-correlation between different $z_s$ maps. Our maps are also useful for constructing the off-diagonal components of covariance matrices of all the auto- and cross-correlations. 

\section{Summary}\label{sec:summary}

We release a large suite of cosmological massive  neutrino simulations, ``MassiveNuS'', including 100 massive neutrino models + 1 massless model. We correctly capture the background expansion as neutrinos turn from relativistic to non-relativistic, as well as the growth of neutrino clustering in response to the nonlinear matter growth. We include 3 varying parameters: the neutrino mass sum $M_\nu$ (ranging from 0 to 0.6~eV), matter density $\Omega_m$, and primordial power spectrum amplitude $A_s$.

Our simulations have continuous outputs of snapshots between $z=45$ and $z=0$ (every 126~comoving Mpc/$h$). Our data products include: 
\begin{enumerate}
\itemsep-0.05in
\item 67 snapshots {\red (1024$^3$ particles, 512 Mpc/$h$)} each for the two fiducial models~(section~\ref{sec:sims}), with position and velocity information;
\item halo catalogues, {\red including $\approx 65$ files for each of the 101 models and around three million halos at $z$=0}, complete down to minimal halo mass $\approx10^{11.5}M_\odot$~(section~\ref{sec:halo});
\item merger trees {\red for each of the 101 models}~(section~\ref{sec:halo});
\item lensing convergence maps (12.25 deg$^2$, {\red 0.1 arcmin resolution}) for four source redshifts $z_s = 1$, 1.5, 2, 2.5~(section~\ref{sec:lens});
\item lensing convergence maps (12.25 deg$^2$, {\red 0.1 arcmin resolution}) for the CMB~(section~\ref{sec:lens}).
\end{enumerate}

Our simulations can be used for a wide range of studies related to massive neutrinos, including the  growth of structure, galaxy formation history, weak lensing tomography, cross-correlations, covariance matrices, and non-Gaussian statistics, just to name a few. {\red The simulation products are made publicly available at \url{http://columbialensing.org}}  through the Skies \& Universe Project\footnote{\url{http://skiesanduniverses.org}}.

\begin{table}
\small
\begin{tabular}{l|c|c|c|c} 
\hline
 Model & $M_\nu$ (eV) & $\Omega_m$ &  $10^9A_s$ & $\sigma_8$\\
\hline
1a	&	0.00000	&	0.3000	&	2.1000	&	0.8523	\\
1b	&	0.10000	&	0.3000	&	2.1000	&	0.8295	\\
2	&	0.06271	&	0.3815	&	2.2004	&	1.0135	\\
3	&	0.06522	&	0.2821	&	1.8826	&	0.7563	\\
4	&	0.06773	&	0.4159	&	1.6231	&	0.9171	\\
5	&	0.07024	&	0.2023	&	2.3075	&	0.6231	\\
6	&	0.07275	&	0.3283	&	2.2883	&	0.9324	\\
7	&	0.07526	&	0.3355	&	1.5659	&	0.7828	\\
8	&	0.07778	&	0.2597	&	2.4333	&	0.8008	\\
9	&	0.08030	&	0.2783	&	2.3824	&	0.8382	\\
10	&	0.08282	&	0.2758	&	1.8292	&	0.7285	\\
11	&	0.08535	&	0.3132	&	1.9913	&	0.8378	\\
12	&	0.08788	&	0.2926	&	1.7376	&	0.7429	\\
13	&	0.09041	&	0.2845	&	2.1757	&	0.8126	\\
14	&	0.09295	&	0.3155	&	2.4949	&	0.9411	\\
15	&	0.09550	&	0.3590	&	2.4624	&	1.0231	\\
16	&	0.09805	&	0.3006	&	1.9744	&	0.8059	\\
17	&	0.10061	&	0.2796	&	2.0161	&	0.7690	\\
18	&	0.10318	&	0.3977	&	2.2607	&	1.0456	\\
19	&	0.10575	&	0.3051	&	1.9117	&	0.8004	\\
20	&	0.10833	&	0.2833	&	2.3595	&	0.8385	\\
21	&	0.11092	&	0.3763	&	2.0404	&	0.9574	\\
22	&	0.11351	&	0.1841	&	1.7932	&	0.4885	\\
23	&	0.11612	&	0.3085	&	1.8056	&	0.7821	\\
24	&	0.11874	&	0.3143	&	2.0079	&	0.8358	\\
25	&	0.12136	&	0.3457	&	1.9483	&	0.8811	\\
26	&	0.12400	&	0.3028	&	2.3174	&	0.8714	\\
27	&	0.12665	&	0.2704	&	1.9658	&	0.7324	\\
28	&	0.12931	&	0.2983	&	2.2342	&	0.8445	\\
29	&	0.13198	&	0.2543	&	1.2886	&	0.5618	\\
30	&	0.13467	&	0.3438	&	2.0324	&	0.8934	\\
31	&	0.13737	&	0.3564	&	1.6678	&	0.8292	\\
32	&	0.14008	&	0.3040	&	2.1198	&	0.8318	\\
33	&	0.14281	&	0.3167	&	2.1596	&	0.8656	\\
34	&	0.14556	&	0.2689	&	2.1839	&	0.7635	\\
35	&	0.14832	&	0.2613	&	1.8724	&	0.6891	\\
36	&	0.15110	&	0.3518	&	2.2429	&	0.9494	\\
37	&	0.15389	&	0.3120	&	2.0484	&	0.8307	\\
38	&	0.15671	&	0.2579	&	2.1356	&	0.7256	\\
39	&	0.15954	&	0.2319	&	1.8620	&	0.6145	\\
40	&	0.16240	&	0.3311	&	2.1039	&	0.8779	\\
41	&	0.16527	&	0.3062	&	2.2171	&	0.8490	\\
42	&	0.16817	&	0.2675	&	1.9211	&	0.7072	\\
43	&	0.17109	&	0.3074	&	2.4475	&	0.8931	\\
44	&	0.17404	&	0.3204	&	2.0564	&	0.8444	\\
45	&	0.17701	&	0.2524	&	2.0644	&	0.6944	\\
46	&	0.18001	&	0.2994	&	1.6465	&	0.7156	\\
47	&	0.18303	&	0.3387	&	2.0243	&	0.8705	\\
48	&	0.18608	&	0.2717	&	2.5322	&	0.8178	\\
49	&	0.18917	&	0.3371	&	2.5769	&	0.9770	\\
\hline
\end{tabular}
\quad
\begin{tabular}{l|c|c|c|c} 
\hline
50	&	0.19228	&	0.2562	&	2.2517	&	0.7308	\\
51	&	0.19543	&	0.3242	&	1.9022	&	0.8143	\\
52	&	0.19861	&	0.2281	&	1.6872	&	0.5664	\\
53	&	0.20183	&	0.2460	&	2.5535	&	0.7474	\\
54	&	0.20508	&	0.2949	&	2.0961	&	0.7912	\\
55	&	0.20837	&	0.2938	&	2.2978	&	0.8250	\\
56	&	0.21171	&	0.3192	&	2.9114	&	0.9905	\\
57	&	0.21508	&	0.2770	&	1.5292	&	0.6394	\\
58	&	0.21851	&	0.2903	&	1.9303	&	0.7464	\\
59	&	0.22198	&	0.3230	&	1.9393	&	0.8133	\\
60	&	0.22550	&	0.3296	&	2.4068	&	0.9192	\\
61	&	0.22907	&	0.3648	&	2.3276	&	0.9720	\\
62	&	0.23270	&	0.3476	&	1.4826	&	0.7488	\\
63	&	0.23638	&	0.3882	&	2.1921	&	0.9827	\\
64	&	0.24013	&	0.2482	&	2.3944	&	0.7188	\\
65	&	0.24394	&	0.2880	&	2.3708	&	0.8523	\\
66	&	0.24782	&	0.3179	&	2.2789	&	0.8643	\\
67	&	0.25177	&	0.2660	&	1.9571	&	0.6891	\\
68	&	0.25580	&	0.2436	&	2.1436	&	0.6641	\\
69	&	0.25991	&	0.2960	&	2.6708	&	0.8800	\\
70	&	0.26411	&	0.3421	&	2.4781	&	0.9486	\\
71	&	0.26840	&	0.3017	&	2.1118	&	0.7927	\\
72	&	0.27278	&	0.2915	&	1.7219	&	0.6946	\\
73	&	0.27728	&	0.3256	&	2.0802	&	0.8337	\\
74	&	0.28188	&	0.2744	&	2.2087	&	0.7446	\\
75	&	0.28661	&	0.2629	&	1.8405	&	0.6528	\\
76	&	0.29147	&	0.3108	&	2.2697	&	0.8359	\\
77	&	0.29647	&	0.2352	&	1.8176	&	0.5813	\\
78	&	0.30162	&	0.2185	&	2.4198	&	0.6205	\\
79	&	0.30694	&	0.3681	&	2.5128	&	0.9964	\\
80	&	0.31244	&	0.2118	&	2.6341	&	0.6234	\\
81	&	0.31814	&	0.2504	&	2.0882	&	0.6564	\\
82	&	0.32406	&	0.3340	&	1.7525	&	0.7701	\\
83	&	0.33021	&	0.3325	&	1.9996	&	0.8184	\\
84	&	0.33664	&	0.2892	&	2.1277	&	0.7508	\\
85	&	0.34337	&	0.2410	&	2.2256	&	0.6462	\\
86	&	0.35044	&	0.2645	&	2.1676	&	0.6958	\\
87	&	0.35789	&	0.3496	&	1.7802	&	0.7965	\\
88	&	0.36578	&	0.2808	&	1.7051	&	0.6483	\\
89	&	0.37417	&	0.2972	&	2.7836	&	0.8684	\\
90	&	0.38317	&	0.2857	&	1.4164	&	0.5964	\\
91	&	0.39287	&	0.2731	&	1.8925	&	0.6595	\\
92	&	0.40343	&	0.3540	&	1.5966	&	0.7521	\\
93	&	0.41504	&	0.3097	&	2.3486	&	0.8157	\\
94	&	0.42800	&	0.2868	&	2.0723	&	0.7132	\\
95	&	0.44273	&	0.3269	&	2.6034	&	0.8913	\\
96	&	0.45988	&	0.2237	&	1.7667	&	0.5066	\\
97	&	0.48062	&	0.3403	&	2.1516	&	0.8287	\\
98	&	0.50723	&	0.2383	&	2.7174	&	0.6595	\\
99	&	0.54543	&	0.3617	&	2.3380	&	0.8916	\\
100	&	0.62036	&	0.3217	&	1.9829	&	0.7283	\\
\hline
\end{tabular}
\caption[]{\label{tab: CosmoParsm} Cosmological parameters for our simulations. {\red A flat universe ($\Omega_\Lambda+\Omega_m=1$) is assumed.} Other parameters are fixed at $h=0.7$, $n_s=0.97$, $\Omega_b=0.046$, and  $w=-1$. All points are visualized in Fig.~\ref{fig:design}.}
\end{table}

\begin{acknowledgments}
We thank Yacine Ali-Ha\"imoud, Nick Battaglia, Marcel Schmittfull, Francisco Villaescusa-Navarro, and Eli Visbal for useful discussions.
JL is supported by an NSF Astronomy and Astrophysics Postdoctoral Fellowship under award AST-1602663. 
SB was supported by NASA through Einstein Postdoctoral Fellowship Award Number PF5-160133. JCH is supported by the Friends of the Institute for Advanced Study. This work is partially supported by NSF grant AST-1210877 (to ZH) and by a ROADS award at Columbia University. {\red The Flatiron Institute is supported by the Simons Foundation.
We thank New Mexico State University (USA) and Instituto de Astrofisica de Andalucia CSIC (Spain) for hosting the Skies \& Universes site for cosmological simulation products.} 
This work used the Extreme Science and Engineering Discovery Environment (XSEDE), which is supported by NSF grant ACI-1053575. The analysis is in part performed at the TIGRESS high performance computer center at Princeton University. 
\end{acknowledgments}

\bibliographystyle{JHEP}
\bibliography{paper}

\providecommand{\href}[2]{#2}\begingroup\raggedright\begin{thebibliography}{10}

\bibitem{Becker-Szendy1992}
R.~{Becker-Szendy}, C.~B. {Bratton}, D.~{Casper}, S.~T. {Dye}, W.~{Gajewski},
  M.~{Goldhaber} et~al., \emph{{Electron- and muon-neutrino content of the
  atmospheric flux}},
  \href{https://doi.org/10.1103/PhysRevD.46.3720}{\emph{\prd} {\bfseries 46}
  (Nov., 1992) 3720--3724}.

\bibitem{Fukuda1998}
Y.~{Fukuda}, T.~{Hayakawa}, E.~{Ichihara}, K.~{Inoue}, K.~{Ishihara},
  H.~{Ishino} et~al., \emph{{Measurements of the Solar Neutrino Flux from
  Super-Kamiokande's First 300 Days}},
  \href{https://doi.org/10.1103/PhysRevLett.81.1158}{\emph{Physical Review
  Letters} {\bfseries 81} (Aug., 1998) 1158--1162},
  [\href{https://arxiv.org/abs/hep-ex/9805021}{{\ttfamily hep-ex/9805021}}].

\bibitem{Ahmed2004}
S.~N. {Ahmed}, A.~E. {Anthony}, E.~W. {Beier}, A.~{Bellerive}, S.~D. {Biller},
  J.~{Boger} et~al., \emph{{Measurement of the Total Active $^{8}$B Solar
  Neutrino Flux at the Sudbury Neutrino Observatory with Enhanced Neutral
  Current Sensitivity}},
  \href{https://doi.org/10.1103/PhysRevLett.92.181301}{\emph{Physical Review
  Letters} {\bfseries 92} (May, 2004) 181301},
  [\href{https://arxiv.org/abs/nucl-ex/0309004}{{\ttfamily nucl-ex/0309004}}].

\bibitem{Olive:2016xmw}
{\scshape Particle Data Group} collaboration, C.~Patrignani et~al.,
  \emph{{Review of Particle Physics}},
  \href{https://doi.org/10.1088/1674-1137/40/10/100001}{\emph{Chin. Phys.}
  {\bfseries C40} (2016) 100001}.

\bibitem{Wolf2010}
J.~{Wolf} and {Katrin Collaboration}, \emph{{The KATRIN neutrino mass
  experiment}}, \href{https://doi.org/10.1016/j.nima.2010.03.030}{\emph{Nuclear
  Instruments and Methods in Physics Research A} {\bfseries 623} (Nov., 2010)
  442--444}, [\href{https://arxiv.org/abs/0810.3281}{{\ttfamily 0810.3281}}].

\bibitem{Esfahani2017}
A.~{Ashtari Esfahani}, D.~M. {Asner}, S.~{B{\"o}ser}, R.~{Cervantes},
  C.~{Claessens}, L.~{de Viveiros} et~al., \emph{{Determining the neutrino mass
  with cyclotron radiation emission spectroscopy---Project 8}},
  \href{https://doi.org/10.1088/1361-6471/aa5b4f}{\emph{Journal of Physics G
  Nuclear Physics} {\bfseries 44} (May, 2017) 054004},
  [\href{https://arxiv.org/abs/1703.02037}{{\ttfamily 1703.02037}}].

\bibitem{LesgourguesPastor2006}
J.~{Lesgourgues} and S.~{Pastor}, \emph{{Massive neutrinos and cosmology}},
  \href{https://doi.org/10.1016/j.physrep.2006.04.001}{\emph{\physrep}
  {\bfseries 429} (July, 2006) 307--379},
  [\href{https://arxiv.org/abs/astro-ph/0603494}{{\ttfamily
  astro-ph/0603494}}].

\bibitem{Wong2011}
Y.~Y.~Y. {Wong}, \emph{{Neutrino Mass in Cosmology: Status and Prospects}},
  \href{https://doi.org/10.1146/annurev-nucl-102010-130252}{\emph{Annual Review
  of Nuclear and Particle Science} {\bfseries 61} (Nov., 2011) 69--98},
  [\href{https://arxiv.org/abs/1111.1436}{{\ttfamily 1111.1436}}].

\bibitem{planck2015xiii}
{Planck Collaboration}, P.~A.~R. {Ade}, N.~{Aghanim}, M.~{Arnaud},
  M.~{Ashdown}, J.~{Aumont} et~al., \emph{{Planck 2015 results. XIII.
  Cosmological parameters}}, {\emph{ArXiv e-prints} (Feb., 2015) },
  [\href{https://arxiv.org/abs/1502.01589}{{\ttfamily 1502.01589}}].

\bibitem{Betoule2014}
M.~{Betoule}, R.~{Kessler}, J.~{Guy}, J.~{Mosher}, D.~{Hardin}, R.~{Biswas}
  et~al., \emph{{Improved cosmological constraints from a joint analysis of the
  SDSS-II and SNLS supernova samples}},
  \href{https://doi.org/10.1051/0004-6361/201423413}{\emph{\aap} {\bfseries
  568} (Aug., 2014) A22}, [\href{https://arxiv.org/abs/1401.4064}{{\ttfamily
  1401.4064}}].

\bibitem{Beutler2011}
F.~{Beutler}, C.~{Blake}, M.~{Colless}, D.~H. {Jones}, L.~{Staveley-Smith},
  L.~{Campbell} et~al., \emph{{The 6dF Galaxy Survey: baryon acoustic
  oscillations and the local Hubble constant}},
  \href{https://doi.org/10.1111/j.1365-2966.2011.19250.x}{\emph{\mnras}
  {\bfseries 416} (Oct., 2011) 3017--3032},
  [\href{https://arxiv.org/abs/1106.3366}{{\ttfamily 1106.3366}}].

\bibitem{Anderson2014}
L.~{Anderson}, {\'E}.~{Aubourg}, S.~{Bailey}, F.~{Beutler}, V.~{Bhardwaj},
  M.~{Blanton} et~al., \emph{{The clustering of galaxies in the SDSS-III Baryon
  Oscillation Spectroscopic Survey: baryon acoustic oscillations in the Data
  Releases 10 and 11 Galaxy samples}},
  \href{https://doi.org/10.1093/mnras/stu523}{\emph{\mnras} {\bfseries 441}
  (June, 2014) 24--62}, [\href{https://arxiv.org/abs/1312.4877}{{\ttfamily
  1312.4877}}].

\bibitem{Ross2015}
A.~J. {Ross}, L.~{Samushia}, C.~{Howlett}, W.~J. {Percival}, A.~{Burden} and
  M.~{Manera}, \emph{{The clustering of the SDSS DR7 main Galaxy sample - I. A
  4 per cent distance measure at z = 0.15}},
  \href{https://doi.org/10.1093/mnras/stv154}{\emph{\mnras} {\bfseries 449}
  (May, 2015) 835--847}, [\href{https://arxiv.org/abs/1409.3242}{{\ttfamily
  1409.3242}}].

\bibitem{Palanque-Delabrouille2015}
N.~{Palanque-Delabrouille}, C.~{Y{\`e}che}, J.~{Lesgourgues}, G.~{Rossi},
  A.~{Borde}, M.~{Viel} et~al., \emph{{Constraint on neutrino masses from
  SDSS-III/BOSS Ly{$\alpha$} forest and other cosmological probes}},
  \href{https://doi.org/10.1088/1475-7516/2015/02/045}{\emph{\jcap} {\bfseries
  2} (Feb., 2015) 045}, [\href{https://arxiv.org/abs/1410.7244}{{\ttfamily
  1410.7244}}].

\bibitem{planck2013xvi}
{Planck Collaboration}, P.~A.~R. {Ade}, N.~{Aghanim}, C.~{Armitage-Caplan},
  M.~{Arnaud}, M.~{Ashdown} et~al., \emph{{Planck 2013 results. XVI.
  Cosmological parameters}},
  \href{https://doi.org/10.1051/0004-6361/201321591}{\emph{\aap} {\bfseries
  571} (Nov., 2014) A16}, [\href{https://arxiv.org/abs/1303.5076}{{\ttfamily
  1303.5076}}].

\bibitem{Bond1980}
J.~R. {Bond}, G.~{Efstathiou} and J.~{Silk}, \emph{{Massive neutrinos and the
  large-scale structure of the universe}},
  \href{https://doi.org/10.1103/PhysRevLett.45.1980}{\emph{Physical Review
  Letters} {\bfseries 45} (Dec., 1980) 1980--1984}.

\bibitem{Zeldovich1980}
Y.~B. {Zeldovich}, A.~A. {Klypin}, M.~Y. {Khlopov} and V.~M.
  {Chechetkin}{\emph{Sov. J. Nucl. Phys.} (1980) 664}.

\bibitem{Ma1994}
C.-P. {Ma} and E.~{Bertschinger}, \emph{{A calculation of the full neutrino
  phase space in cold + hot dark matter models}},
  \href{https://doi.org/10.1086/174298}{\emph{\apj} {\bfseries 429} (July,
  1994) 22--28}, [\href{https://arxiv.org/abs/astro-ph/9308006}{{\ttfamily
  astro-ph/9308006}}].

\bibitem{Valdarnini1998}
R.~{Valdarnini}, T.~{Kahniashvili} and B.~{Novosyadlyj}, \emph{{Large scale
  structure formation in mixed dark matter models with a cosmological
  constant}}, {\emph{\aap} {\bfseries 336} (Aug., 1998) 11--28},
  [\href{https://arxiv.org/abs/astro-ph/9804057}{{\ttfamily
  astro-ph/9804057}}].

\bibitem{smith2003}
R.~E. {Smith}, J.~A. {Peacock}, A.~{Jenkins}, S.~D.~M. {White}, C.~S. {Frenk},
  F.~R. {Pearce} et~al., \emph{{Stable clustering, the halo model and
  non-linear cosmological power spectra}},
  \href{https://doi.org/10.1046/j.1365-8711.2003.06503.x}{\emph{\mnras}
  {\bfseries 341} (June, 2003) 1311--1332},
  [\href{https://arxiv.org/abs/arXiv:astro-ph/0207664}{{\ttfamily
  arXiv:astro-ph/0207664}}].

\bibitem{takahashi2012}
R.~{Takahashi}, M.~{Sato}, T.~{Nishimichi}, A.~{Taruya} and M.~{Oguri},
  \emph{{Revising the Halofit Model for the Nonlinear Matter Power Spectrum}},
  \href{https://doi.org/10.1088/0004-637X/761/2/152}{\emph{\apj} {\bfseries
  761} (Dec., 2012) 152}, [\href{https://arxiv.org/abs/1208.2701}{{\ttfamily
  1208.2701}}].

\bibitem{Bird2012}
S.~{Bird}, M.~{Viel} and M.~G. {Haehnelt}, \emph{{Massive neutrinos and the
  non-linear matter power spectrum}},
  \href{https://doi.org/10.1111/j.1365-2966.2011.20222.x}{\emph{\mnras}
  {\bfseries 420} (Mar., 2012) 2551--2561},
  [\href{https://arxiv.org/abs/1109.4416}{{\ttfamily 1109.4416}}].

\bibitem{Mead2015}
A.~J. {Mead}, J.~A. {Peacock}, C.~{Heymans}, S.~{Joudaki} and A.~F. {Heavens},
  \emph{{An accurate halo model for fitting non-linear cosmological power
  spectra and baryonic feedback models}},
  \href{https://doi.org/10.1093/mnras/stv2036}{\emph{\mnras} {\bfseries 454}
  (Dec., 2015) 1958--1975}, [\href{https://arxiv.org/abs/1505.07833}{{\ttfamily
  1505.07833}}].

\bibitem{Mead2016}
A.~J. {Mead}, C.~{Heymans}, L.~{Lombriser}, J.~A. {Peacock}, O.~I. {Steele} and
  H.~A. {Winther}, \emph{{Accurate halo-model matter power spectra with dark
  energy, massive neutrinos and modified gravitational forces}},
  \href{https://doi.org/10.1093/mnras/stw681}{\emph{\mnras} {\bfseries 459}
  (June, 2016) 1468--1488}, [\href{https://arxiv.org/abs/1602.02154}{{\ttfamily
  1602.02154}}].

\bibitem{Casarini2009}
L.~{Casarini}, A.~V. {Macci{\`o}} and S.~A. {Bonometto}, \emph{{Dynamical dark
  energy simulations: high accuracy power spectra at high redshift}},
  \href{https://doi.org/10.1088/1475-7516/2009/03/014}{\emph{\jcap} {\bfseries
  3} (Mar., 2009) 014}, [\href{https://arxiv.org/abs/0810.0190}{{\ttfamily
  0810.0190}}].

\bibitem{Casarini2016}
L.~{Casarini}, S.~A. {Bonometto}, E.~{Tessarotto} and P.-S. {Corasaniti},
  \emph{{Extending the Coyote emulator to dark energy models with standard
  w$_{0}$-w$_{a}$ parametrization of the equation of state}},
  \href{https://doi.org/10.1088/1475-7516/2016/08/008}{\emph{\jcap} {\bfseries
  8} (Aug., 2016) 008}, [\href{https://arxiv.org/abs/1601.07230}{{\ttfamily
  1601.07230}}].

\bibitem{PeacockSmith2000}
J.~A. {Peacock} and R.~E. {Smith}, \emph{{Halo occupation numbers and galaxy
  bias}}, \href{https://doi.org/10.1046/j.1365-8711.2000.03779.x}{\emph{\mnras}
  {\bfseries 318} (Nov., 2000) 1144--1156},
  [\href{https://arxiv.org/abs/astro-ph/0005010}{{\ttfamily
  astro-ph/0005010}}].

\bibitem{Seljak2000}
U.~{Seljak}, \emph{{Analytic model for galaxy and dark matter clustering}},
  \href{https://doi.org/10.1046/j.1365-8711.2000.03715.x}{\emph{\mnras}
  {\bfseries 318} (Oct., 2000) 203--213},
  [\href{https://arxiv.org/abs/astro-ph/0001493}{{\ttfamily
  astro-ph/0001493}}].

\bibitem{heitmann2010i}
K.~{Heitmann}, M.~{White}, C.~{Wagner}, S.~{Habib} and D.~{Higdon}, \emph{{The
  Coyote Universe. I. Precision Determination of the Nonlinear Matter Power
  Spectrum}}, \href{https://doi.org/10.1088/0004-637X/715/1/104}{\emph{\apj}
  {\bfseries 715} (May, 2010) 104--121},
  [\href{https://arxiv.org/abs/0812.1052}{{\ttfamily 0812.1052}}].

\bibitem{heitmann2009ii}
K.~{Heitmann}, D.~{Higdon}, M.~{White}, S.~{Habib}, B.~J. {Williams},
  E.~{Lawrence} et~al., \emph{{The Coyote Universe. II. Cosmological Models and
  Precision Emulation of the Nonlinear Matter Power Spectrum}},
  \href{https://doi.org/10.1088/0004-637X/705/1/156}{\emph{\apj} {\bfseries
  705} (Nov., 2009) 156--174},
  [\href{https://arxiv.org/abs/0902.0429}{{\ttfamily 0902.0429}}].

\bibitem{heitmann2010iii}
E.~{Lawrence}, K.~{Heitmann}, M.~{White}, D.~{Higdon}, C.~{Wagner}, S.~{Habib}
  et~al., \emph{{The Coyote Universe. III. Simulation Suite and Precision
  Emulator for the Nonlinear Matter Power Spectrum}},
  \href{https://doi.org/10.1088/0004-637X/713/2/1322}{\emph{\apj} {\bfseries
  713} (Apr., 2010) 1322--1331},
  [\href{https://arxiv.org/abs/0912.4490}{{\ttfamily 0912.4490}}].

\bibitem{heitmann2014e}
K.~{Heitmann}, E.~{Lawrence}, J.~{Kwan}, S.~{Habib} and D.~{Higdon}, \emph{{The
  Coyote Universe Extended: Precision Emulation of the Matter Power Spectrum}},
  \href{https://doi.org/10.1088/0004-637X/780/1/111}{\emph{\apj} {\bfseries
  780} (Jan., 2014) 111}, [\href{https://arxiv.org/abs/1304.7849}{{\ttfamily
  1304.7849}}].

\bibitem{AlihaimoundBird2013}
Y.~{Ali-Ha{\"i}moud} and S.~{Bird}, \emph{{An efficient implementation of
  massive neutrinos in non-linear structure formation simulations}},
  \href{https://doi.org/10.1093/mnras/sts286}{\emph{\mnras} {\bfseries 428}
  (Feb., 2013) 3375--3389}, [\href{https://arxiv.org/abs/1209.0461}{{\ttfamily
  1209.0461}}].

\bibitem{Bird2017}
S.~{Bird} and et~al.

\bibitem{Brandbyge2009}
J.~{Brandbyge} and S.~{Hannestad}, \emph{{Grid based linear neutrino
  perturbations in cosmological N-body simulations}},
  \href{https://doi.org/10.1088/1475-7516/2009/05/002}{\emph{\jcap} {\bfseries
  5} (May, 2009) 002}, [\href{https://arxiv.org/abs/0812.3149}{{\ttfamily
  0812.3149}}].

\bibitem{Viel2010}
M.~{Viel}, M.~G. {Haehnelt} and V.~{Springel}, \emph{{The effect of neutrinos
  on the matter distribution as probed by the intergalactic medium}},
  \href{https://doi.org/10.1088/1475-7516/2010/06/015}{\emph{\jcap} {\bfseries
  6} (June, 2010) 015}, [\href{https://arxiv.org/abs/1003.2422}{{\ttfamily
  1003.2422}}].

\bibitem{Petri2016Lenstools}
A.~{Petri}, \emph{{Mocking the weak lensing universe: The LensTools Python
  computing package}},
  \href{https://doi.org/10.1016/j.ascom.2016.06.001}{\emph{Astronomy and
  Computing} {\bfseries 17} (Oct., 2016) 73--79},
  [\href{https://arxiv.org/abs/1606.01903}{{\ttfamily 1606.01903}}].

\bibitem{Brandbyge2008}
J.~{Brandbyge}, S.~{Hannestad}, T.~{Haugb{\o}lle} and B.~{Thomsen}, \emph{{The
  effect of thermal neutrino motion on the non-linear cosmological matter power
  spectrum}}, \href{https://doi.org/10.1088/1475-7516/2008/08/020}{\emph{\jcap}
  {\bfseries 8} (Aug., 2008) 020},
  [\href{https://arxiv.org/abs/0802.3700}{{\ttfamily 0802.3700}}].

\bibitem{Villaescusa-Navarro2013}
F.~{Villaescusa-Navarro}, S.~{Bird}, C.~{Pe{\~n}a-Garay} and M.~{Viel},
  \emph{{Non-linear evolution of the cosmic neutrino background}},
  \href{https://doi.org/10.1088/1475-7516/2013/03/019}{\emph{\jcap} {\bfseries
  3} (Mar., 2013) 019}, [\href{https://arxiv.org/abs/1212.4855}{{\ttfamily
  1212.4855}}].

\bibitem{Castorina2015}
E.~{Castorina}, C.~{Carbone}, J.~{Bel}, E.~{Sefusatti} and K.~{Dolag},
  \emph{{DEMNUni: the clustering of large-scale structures in the presence of
  massive neutrinos}},
  \href{https://doi.org/10.1088/1475-7516/2015/07/043}{\emph{\jcap} {\bfseries
  7} (July, 2015) 043}, [\href{https://arxiv.org/abs/1505.07148}{{\ttfamily
  1505.07148}}].

\bibitem{Adamek2017}
J.~{Adamek}, R.~{Durrer} and M.~{Kunz}, \emph{{Relativistic N-body simulations
  with massive neutrinos}}, {\emph{ArXiv e-prints} (July, 2017) },
  [\href{https://arxiv.org/abs/1707.06938}{{\ttfamily 1707.06938}}].

\bibitem{Emberson2017}
J.~D. {Emberson}, H.-R. {Yu}, D.~{Inman}, T.-J. {Zhang}, U.-L. {Pen},
  J.~{Harnois-D{\'e}raps} et~al., \emph{{Cosmological neutrino simulations at
  extreme scale}},
  \href{https://doi.org/10.1088/1674-4527/17/8/85}{\emph{Research in Astronomy
  and Astrophysics} {\bfseries 17} (Aug., 2017) 085},
  [\href{https://arxiv.org/abs/1611.01545}{{\ttfamily 1611.01545}}].

\bibitem{Wagner2012}
C.~{Wagner}, L.~{Verde} and R.~{Jimenez}, \emph{{Effects of the Neutrino Mass
  Splitting on the Nonlinear Matter Power Spectrum}},
  \href{https://doi.org/10.1088/2041-8205/752/2/L31}{\emph{\apjl} {\bfseries
  752} (June, 2012) L31}, [\href{https://arxiv.org/abs/1203.5342}{{\ttfamily
  1203.5342}}].

\bibitem{Brandbyge2010}
J.~{Brandbyge} and S.~{Hannestad}, \emph{{Resolving cosmic neutrino structure:
  a hybrid neutrino N-body scheme}},
  \href{https://doi.org/10.1088/1475-7516/2010/01/021}{\emph{\jcap} {\bfseries
  1} (Jan., 2010) 021}, [\href{https://arxiv.org/abs/0908.1969}{{\ttfamily
  0908.1969}}].

\bibitem{Inman2017dip}
D.~{Inman}, H.-R. {Yu}, H.-M. {Zhu}, J.~D. {Emberson}, U.-L. {Pen}, T.-J.
  {Zhang} et~al., \emph{{Simulating the cold dark matter-neutrino dipole with
  TianNu}}, \href{https://doi.org/10.1103/PhysRevD.95.083518}{\emph{\prd}
  {\bfseries 95} (Apr., 2017) 083518}.

\bibitem{Inman2017}
D.~{Inman} and U.-L. {Pen}, \emph{{Cosmic neutrinos: A dispersive and nonlinear
  fluid}}, \href{https://doi.org/10.1103/PhysRevD.95.063535}{\emph{\prd}
  {\bfseries 95} (Mar., 2017) 063535},
  [\href{https://arxiv.org/abs/1609.09469}{{\ttfamily 1609.09469}}].

\bibitem{Banerjee2016}
A.~{Banerjee} and N.~{Dalal}, \emph{{Simulating nonlinear cosmological
  structure formation with massive neutrinos}},
  \href{https://doi.org/10.1088/1475-7516/2016/11/015}{\emph{\jcap} {\bfseries
  11} (Nov., 2016) 015}, [\href{https://arxiv.org/abs/1606.06167}{{\ttfamily
  1606.06167}}].

\bibitem{springel2005}
V.~{Springel}, \emph{{The cosmological simulation code GADGET-2}},
  \href{https://doi.org/10.1111/j.1365-2966.2005.09655.x}{\emph{\mnras}
  {\bfseries 364} (Dec., 2005) 1105--1134},
  [\href{https://arxiv.org/abs/arXiv:astro-ph/0505010}{{\ttfamily
  arXiv:astro-ph/0505010}}].

\bibitem{Zeldovich1970}
Y.~B. {Zel'dovich}, \emph{{Gravitational instability: An approximate theory for
  large density perturbations.}}, {\emph{\aap} {\bfseries 5} (Mar., 1970)
  84--89}.

\bibitem{MaBertschinger1994}
C.-P. {Ma} and E.~{Bertschinger}, \emph{{Cosmological Perturbation Theory in
  the Synchronous vs. Conformal Newtonian Gauge}}, {\emph{ArXiv Astrophysics
  e-prints} (Jan., 1994) },
  [\href{https://arxiv.org/abs/astro-ph/9401007}{{\ttfamily
  astro-ph/9401007}}].

\bibitem{Agarwal2011}
S.~{Agarwal} and H.~A. {Feldman}, \emph{{The effect of massive neutrinos on the
  matter power spectrum}},
  \href{https://doi.org/10.1111/j.1365-2966.2010.17546.x}{\emph{\mnras}
  {\bfseries 410} (Jan., 2011) 1647--1654},
  [\href{https://arxiv.org/abs/1006.0689}{{\ttfamily 1006.0689}}].

\bibitem{Massara2014}
E.~{Massara}, F.~{Villaescusa-Navarro} and M.~{Viel}, \emph{{The halo model in
  a massive neutrino cosmology}},
  \href{https://doi.org/10.1088/1475-7516/2014/12/053}{\emph{\jcap} {\bfseries
  12} (Dec., 2014) 053}, [\href{https://arxiv.org/abs/1410.6813}{{\ttfamily
  1410.6813}}].

\bibitem{cooray2002}
A.~{Cooray} and R.~{Sheth}, \emph{{Halo models of large scale structure}},
  \href{https://doi.org/10.1016/S0370-1573(02)00276-4}{\emph{\physrep}
  {\bfseries 372} (Dec., 2002) 1--129},
  [\href{https://arxiv.org/abs/astro-ph/0206508}{{\ttfamily
  astro-ph/0206508}}].

\bibitem{Behroozi2013Rockstar}
P.~S. {Behroozi}, R.~H. {Wechsler} and H.-Y. {Wu}, \emph{{The ROCKSTAR
  Phase-space Temporal Halo Finder and the Velocity Offsets of Cluster Cores}},
  \href{https://doi.org/10.1088/0004-637X/762/2/109}{\emph{\apj} {\bfseries
  762} (Jan., 2013) 109}, [\href{https://arxiv.org/abs/1110.4372}{{\ttfamily
  1110.4372}}].

\bibitem{Peebles1969}
P.~J.~E. {Peebles}, \emph{{Origin of the Angular Momentum of Galaxies}},
  \href{https://doi.org/10.1086/149876}{\emph{\apj} {\bfseries 155} (Feb.,
  1969) 393}.

\bibitem{Tinker2010}
J.~L. {Tinker}, B.~E. {Robertson}, A.~V. {Kravtsov}, A.~{Klypin}, M.~S.
  {Warren}, G.~{Yepes} et~al., \emph{{The Large-scale Bias of Dark Matter
  Halos: Numerical Calibration and Model Tests}},
  \href{https://doi.org/10.1088/0004-637X/724/2/878}{\emph{\apj} {\bfseries
  724} (Dec., 2010) 878--886},
  [\href{https://arxiv.org/abs/1001.3162}{{\ttfamily 1001.3162}}].

\bibitem{Ichiki2012}
K.~{Ichiki} and M.~{Takada}, \emph{{Impact of massive neutrinos on the
  abundance of massive clusters}},
  \href{https://doi.org/10.1103/PhysRevD.85.063521}{\emph{\prd} {\bfseries 85}
  (Mar., 2012) 063521}, [\href{https://arxiv.org/abs/1108.4688}{{\ttfamily
  1108.4688}}].

\bibitem{Behroozi2013Tree}
P.~S. {Behroozi}, R.~H. {Wechsler}, H.-Y. {Wu}, M.~T. {Busha}, A.~A. {Klypin}
  and J.~R. {Primack}, \emph{{Gravitationally Consistent Halo Catalogs and
  Merger Trees for Precision Cosmology}},
  \href{https://doi.org/10.1088/0004-637X/763/1/18}{\emph{\apj} {\bfseries 763}
  (Jan., 2013) 18}, [\href{https://arxiv.org/abs/1110.4370}{{\ttfamily
  1110.4370}}].

\bibitem{Kilbinger2015}
M.~{Kilbinger}, \emph{{Cosmology with cosmic shear observations: a review}},
  \href{https://doi.org/10.1088/0034-4885/78/8/086901}{\emph{Reports on
  Progress in Physics} {\bfseries 78} (July, 2015) 086901},
  [\href{https://arxiv.org/abs/1411.0115}{{\ttfamily 1411.0115}}].

\bibitem{Hanson2010}
D.~{Hanson}, A.~{Challinor} and A.~{Lewis}, \emph{{Weak lensing of the CMB}},
  \href{https://doi.org/10.1007/s10714-010-1036-y}{\emph{General Relativity and
  Gravitation} {\bfseries 42} (Sept., 2010) 2197--2218},
  [\href{https://arxiv.org/abs/0911.0612}{{\ttfamily 0911.0612}}].

\bibitem{Das2011}
S.~{Das}, B.~D. {Sherwin}, P.~{Aguirre}, J.~W. {Appel}, J.~R. {Bond}, C.~S.
  {Carvalho} et~al., \emph{{Detection of the Power Spectrum of Cosmic Microwave
  Background Lensing by the Atacama Cosmology Telescope}},
  \href{https://doi.org/10.1103/PhysRevLett.107.021301}{\emph{Physical Review
  Letters} {\bfseries 107} (July, 2011) 021301},
  [\href{https://arxiv.org/abs/1103.2124}{{\ttfamily 1103.2124}}].

\bibitem{vanEngelen2012}
A.~{van Engelen}, R.~{Keisler}, O.~{Zahn}, K.~A. {Aird}, B.~A. {Benson}, L.~E.
  {Bleem} et~al., \emph{{A Measurement of Gravitational Lensing of the
  Microwave Background Using South Pole Telescope Data}},
  \href{https://doi.org/10.1088/0004-637X/756/2/142}{\emph{\apj} {\bfseries
  756} (Sept., 2012) 142}, [\href{https://arxiv.org/abs/1202.0546}{{\ttfamily
  1202.0546}}].

\bibitem{Planck2015XV}
{Planck Collaboration}, P.~A.~R. {Ade}, N.~{Aghanim}, M.~{Arnaud},
  M.~{Ashdown}, J.~{Aumont} et~al., \emph{{Planck 2015 results. XV.
  Gravitational lensing}}, {\emph{ArXiv e-prints} (Feb., 2015) },
  [\href{https://arxiv.org/abs/1502.01591}{{\ttfamily 1502.01591}}].

\bibitem{Sherwin2016ACTPol}
B.~D. {Sherwin}, A.~{van Engelen}, N.~{Sehgal}, M.~{Madhavacheril}, G.~E.
  {Addison}, S.~{Aiola} et~al., \emph{{The Atacama Cosmology Telescope:
  Two-Season ACTPol Lensing Power Spectrum}}, {\emph{ArXiv e-prints} (Nov.,
  2016) }, [\href{https://arxiv.org/abs/1611.09753}{{\ttfamily 1611.09753}}].

\bibitem{Omori2017}
Y.~{Omori}, R.~{Chown}, G.~{Simard}, K.~T. {Story}, K.~{Aylor}, E.~J. {Baxter}
  et~al., \emph{{A 2500 square-degree CMB lensing map from combined South Pole
  Telescope and Planck data}}, {\emph{ArXiv e-prints} (May, 2017) },
  [\href{https://arxiv.org/abs/1705.00743}{{\ttfamily 1705.00743}}].

\bibitem{Schrabback2010}
T.~{Schrabback}, J.~{Hartlap}, B.~{Joachimi}, M.~{Kilbinger}, P.~{Simon},
  K.~{Benabed} et~al., \emph{{Evidence of the accelerated expansion of the
  Universe from weak lensing tomography with COSMOS}},
  \href{https://doi.org/10.1051/0004-6361/200913577}{\emph{\aap} {\bfseries
  516} (June, 2010) A63}, [\href{https://arxiv.org/abs/0911.0053}{{\ttfamily
  0911.0053}}].

\bibitem{Heymans2012}
C.~{Heymans}, L.~{Van Waerbeke}, L.~{Miller}, T.~{Erben}, H.~{Hildebrandt},
  H.~{Hoekstra} et~al., \emph{{CFHTLenS: the Canada-France-Hawaii Telescope
  Lensing Survey}},
  \href{https://doi.org/10.1111/j.1365-2966.2012.21952.x}{\emph{\mnras}
  {\bfseries 427} (Nov., 2012) 146--166},
  [\href{https://arxiv.org/abs/1210.0032}{{\ttfamily 1210.0032}}].

\bibitem{Hildebrandt2017}
H.~{Hildebrandt}, M.~{Viola}, C.~{Heymans}, S.~{Joudaki}, K.~{Kuijken},
  C.~{Blake} et~al., \emph{{KiDS-450: cosmological parameter constraints from
  tomographic weak gravitational lensing}},
  \href{https://doi.org/10.1093/mnras/stw2805}{\emph{\mnras} {\bfseries 465}
  (Feb., 2017) 1454--1498}, [\href{https://arxiv.org/abs/1606.05338}{{\ttfamily
  1606.05338}}].

\bibitem{Mandelbaum2017}
R.~{Mandelbaum}, H.~{Miyatake}, T.~{Hamana}, M.~{Oguri}, M.~{Simet},
  R.~{Armstrong} et~al., \emph{{The first-year shear catalog of the Subaru
  Hyper Suprime-Cam SSP Survey}}, {\emph{ArXiv e-prints} (May, 2017) },
  [\href{https://arxiv.org/abs/1705.06745}{{\ttfamily 1705.06745}}].

\bibitem{2017DES}
{DES Collaboration}, T.~M.~C. {Abbott}, F.~B. {Abdalla}, A.~{Alarcon},
  J.~{Aleksi{\'c}}, S.~{Allam} et~al., \emph{{Dark Energy Survey Year 1
  Results: Cosmological Constraints from Galaxy Clustering and Weak Lensing}},
  {\emph{ArXiv e-prints} (Aug., 2017) },
  [\href{https://arxiv.org/abs/1708.01530}{{\ttfamily 1708.01530}}].

\bibitem{Hilbert2009}
S.~{Hilbert}, J.~{Hartlap}, S.~D.~M. {White} and P.~{Schneider},
  \emph{{Ray-tracing through the Millennium Simulation: Born corrections and
  lens-lens coupling in cosmic shear and galaxy-galaxy lensing}},
  \href{https://doi.org/10.1051/0004-6361/200811054}{\emph{\aap} {\bfseries
  499} (May, 2009) 31--43}, [\href{https://arxiv.org/abs/0809.5035}{{\ttfamily
  0809.5035}}].

\bibitem{Schafer2012}
B.~M. {Sch{\"a}fer}, L.~{Heisenberg}, A.~F. {Kalovidouris} and D.~J. {Bacon},
  \emph{{On the validity of the Born approximation for weak cosmic flexions}},
  \href{https://doi.org/10.1111/j.1365-2966.2011.20051.x}{\emph{\mnras}
  {\bfseries 420} (Feb., 2012) 455--467},
  [\href{https://arxiv.org/abs/1101.4769}{{\ttfamily 1101.4769}}].

\bibitem{Pratten2016}
G.~{Pratten} and A.~{Lewis}, \emph{{Impact of post-Born lensing on the CMB}},
  \href{https://doi.org/10.1088/1475-7516/2016/08/047}{\emph{\jcap} {\bfseries
  8} (Aug., 2016) 047}, [\href{https://arxiv.org/abs/1605.05662}{{\ttfamily
  1605.05662}}].

\bibitem{Petri2017}
A.~{Petri}, Z.~{Haiman} and M.~{May}, \emph{{Validity of the Born approximation
  for beyond Gaussian weak lensing observables}},
  \href{https://doi.org/10.1103/PhysRevD.95.123503}{\emph{\prd} {\bfseries 95}
  (June, 2017) 123503}, [\href{https://arxiv.org/abs/1612.00852}{{\ttfamily
  1612.00852}}].

\bibitem{Jain2000}
B.~{Jain}, U.~{Seljak} and S.~{White}, \emph{{Ray-tracing Simulations of Weak
  Lensing by Large-Scale Structure}},
  \href{https://doi.org/10.1086/308384}{\emph{\apj} {\bfseries 530} (Feb.,
  2000) 547--577},
  [\href{https://arxiv.org/abs/arXiv:astro-ph/9901191}{{\ttfamily
  arXiv:astro-ph/9901191}}].

\bibitem{Petri2016a}
A.~{Petri}, Z.~{Haiman} and M.~{May}, \emph{{Sample variance in weak lensing:
  How many simulations are required?}},
  \href{https://doi.org/10.1103/PhysRevD.93.063524}{\emph{\prd} {\bfseries 93}
  (Mar., 2016) 063524}, [\href{https://arxiv.org/abs/1601.06792}{{\ttfamily
  1601.06792}}].

\bibitem{shirasakiyoshida2014}
M.~{Shirasaki} and N.~{Yoshida}, \emph{{Statistical and Systematic Errors in
  the Measurement of Weak-Lensing Minkowski Functionals: Application to the
  Canada-France-Hawaii Lensing Survey}},
  \href{https://doi.org/10.1088/0004-637X/786/1/43}{\emph{\apj} {\bfseries 786}
  (May, 2014) 43}, [\href{https://arxiv.org/abs/1312.5032}{{\ttfamily
  1312.5032}}].

\bibitem{Liu2015}
J.~{Liu}, A.~{Petri}, Z.~{Haiman}, L.~{Hui}, J.~M. {Kratochvil} and M.~{May},
  \emph{{Cosmology constraints from the weak lensing peak counts and the power
  spectrum in CFHTLenS data}},
  \href{https://doi.org/10.1103/PhysRevD.91.063507}{\emph{\prd} {\bfseries 91}
  (Mar., 2015) 063507}, [\href{https://arxiv.org/abs/1412.0757}{{\ttfamily
  1412.0757}}].

\bibitem{Petri2015}
A.~{Petri}, J.~{Liu}, Z.~{Haiman}, M.~{May}, L.~{Hui} and J.~M. {Kratochvil},
  \emph{{Emulating the CFHTLenS weak lensing data: Cosmological constraints
  from moments and Minkowski functionals}},
  \href{https://doi.org/10.1103/PhysRevD.91.103511}{\emph{\prd} {\bfseries 91}
  (May, 2015) 103511}, [\href{https://arxiv.org/abs/1503.06214}{{\ttfamily
  1503.06214}}].

\bibitem{Liux2015}
X.~{Liu}, C.~{Pan}, R.~{Li}, H.~{Shan}, Q.~{Wang}, L.~{Fu} et~al.,
  \emph{{Cosmological constraints from weak lensing peak statistics with
  Canada-France-Hawaii Telescope Stripe 82 Survey}},
  \href{https://doi.org/10.1093/mnras/stv784}{\emph{\mnras} {\bfseries 450}
  (July, 2015) 2888--2902}, [\href{https://arxiv.org/abs/1412.3683}{{\ttfamily
  1412.3683}}].

\bibitem{Kacprzak2016}
T.~{Kacprzak}, D.~{Kirk}, O.~{Friedrich}, A.~{Amara}, A.~{Refregier},
  L.~{Marian} et~al., \emph{{Cosmology constraints from shear peak statistics
  in Dark Energy Survey Science Verification data}},
  \href{https://doi.org/10.1093/mnras/stw2070}{\emph{\mnras} {\bfseries 463}
  (Dec., 2016) 3653--3673}, [\href{https://arxiv.org/abs/1603.05040}{{\ttfamily
  1603.05040}}].

\bibitem{Martinet2017}
N.~{Martinet}, P.~{Schneider}, H.~{Hildebrandt}, H.~{Shan}, M.~{Asgari}, J.~P.
  {Dietrich} et~al., \emph{{KiDS-450: Cosmological Constraints from Weak
  Lensing Peak Statistics - II: Inference from Shear Peaks in N-body
  Simulations}}, {\emph{ArXiv e-prints} (Sept., 2017) },
  [\href{https://arxiv.org/abs/1709.07678}{{\ttfamily 1709.07678}}].

\bibitem{Shan2017}
H.~{Shan}, X.~{Liu}, H.~{Hildebrandt}, C.~{Pan}, N.~{Martinet}, Z.~{Fan}
  et~al., \emph{{KiDS-450: Cosmological Constraints from Weak Lensing Peak
  Statistics-I: Inference from Analytical Prediction of high Signal-to-Noise
  Ratio Convergence Peaks}}, {\emph{ArXiv e-prints} (Sept., 2017) },
  [\href{https://arxiv.org/abs/1709.07651}{{\ttfamily 1709.07651}}].

\bibitem{Namikawa2016}
T.~{Namikawa}, \emph{{CMB lensing bispectrum from nonlinear growth of the large
  scale structure}},
  \href{https://doi.org/10.1103/PhysRevD.93.121301}{\emph{\prd} {\bfseries 93}
  (June, 2016) 121301}, [\href{https://arxiv.org/abs/1604.08578}{{\ttfamily
  1604.08578}}].

\bibitem{Liu2016b}
J.~{Liu}, J.~C. {Hill}, B.~D. {Sherwin}, A.~{Petri}, V.~{B{\"o}hm} and
  Z.~{Haiman}, \emph{{CMB lensing beyond the power spectrum: Cosmological
  constraints from the one-point probability distribution function and peak
  counts}}, \href{https://doi.org/10.1103/PhysRevD.94.103501}{\emph{\prd}
  {\bfseries 94} (Nov., 2016) 103501},
  [\href{https://arxiv.org/abs/1608.03169}{{\ttfamily 1608.03169}}].

\end{thebibliography}\endgroup
\end{document}